\documentclass{aa}  
\usepackage{graphicx}
\usepackage{natbib}

\bibpunct{(}{)}{;}{a}{}{,}   % necessary for bibtex in A&A as well
\usepackage{txfonts}
\begin{document}
\title{Hot ammonia in NGC6334I \& I(N)}

%   \subtitle{I. Overviewing the $\kappa$-mechanism}

   \author{H.~Beuther\inst{1},
          A.J. Walsh\inst{2},
          S. Thorwirth\inst{3},
          Q. Zhang\inst{4},
          T.R. Hunter\inst{5},
          S.T. Megeath\inst{6}
          and
          K.M. Menten\inst{3}
          }

   \offprints{H.~Beuther}

   \institute{Max-Planck-Institut f\"ur Astronomie, K\"onigstuhl 17, 
              69117 Heidelberg, Germany\\
              \email{beuther@mpia.de}
         \and
              James Cook University,
              Townsville, QLD 4811, Australia \\
              \email{Andrew.Walsh@jcu.edu.au}
         \and
              Max-Planck-Institut f\"ur Radioastronomie, Auf dem H\"ugel 69, 
              53121 Bonn, Germany\\
              \email{sthorwirth@mpifr-bonn.mpg.de}
         \and
              Harvard-Smithsonian Center for Astrophysics, 60 Garden Street,
              Cambridge, MA 02138, USA\\
             \email{zhang@cfa.harvard.edu}
         \and
              NRAO, 520 Edgemont Rd,
              Charlottesville, VA 22903\\
              \email{thunter@nrao.edu}
         \and
              Ritter Observatory, Department of Physics and Astronomy, 
              University of Toledo, Toledo, OH 43606-3390, USA\\
              \email{megeath@astro1.panet.utoledo.edu}
%             \thanks{The university of heaven temporarily does not
%                     accept e-mails}
            } 
\authorrunning{Beuther et al.} 
\titlerunning{Hot ammonia from NGC6334I \& I(N)}

   \date{}

% \abstract{}{}{}{}{} 
% 5 {} token are mandatory 
  \abstract
    {}
    {The massive twin cores NGC6334I and I(N) are in different
      evolutionary stages and hence ideal targets to study
      evolutionary variations within the same larger-scale
      environment. Here, we study the warm, compact gas
      components.}
    {We imaged the two regions with the Australia Telescope Compact
      Array (ATCA) at high angular resolution in the NH$_3$(3,3) to
      (6,6) inversion lines.}
    {Compact emission is detected toward both regions in all
      observed inversion lines with energy levels up to 407\,K above
      ground.  This is particularly surprising for NGC6334I(N) since
      it lacks bright infrared emission and is considered a massive
      cold core at an early evolutionary stage. High optical depth and
      multiply-peaked line profiles complicate rotation temperature
      estimates, and we can
      only conclude that gas components with temperatures $>100$\,K
      are present in both regions. Toward NGC6334I, we confirm
      previous reports of NH$_3$(3,3) maser emission toward the
      outflow bow-shocks. Furthermore, we report the first detection
      of an NH$_3$(6,6) maser toward the central region of NGC6334I.
      This maser is centered on the second millimeter (mm) peak and
      elongated along the outflow axis, indicating that this mm
      continuum core harbors the driving source of the molecular
      outflow. Toward the main mm peak in NGC6334I(N), we detect a
      double-horn line profile in the NH$_3$(6,6) transition. The
      current data do not allow us to differentiate whether this
      double-horn profile is produced by multiple gas components along
      the line of sight, or whether it may trace a potential
      underlying massive accretion disk. {{\it The data to Figures 3
          to 7 are also available in electronic form at the CDS via
          anonymous ftp to cdsarc.u-strasbg.fr (130.79.128.5) or via
          http://cdsweb.u-strasbg.fr/cgi-bin/qcat?J/A+A/.}}}
   {}
   

   \maketitle
%
%________________________________________________________________

\section{Introduction}

The massive star-forming regions NGC6334I \& I(N) are located at the
north-eastern end of the NGC6334 molecular cloud/H{\sc ii} region
complex at an approximate distance of 1.7\,kpc in the southern
hemisphere \citep{neckel1978,straw1989}. The whole NGC6334 complex has
been subject to intense studies in many wavelength bands for more
than two decades (e.g.,
\citealt{mcbreen1979,rodriguez1982,gezari1982,loughran1986,depree1995,tapia1996,sandell2000,carral2002,beuther2005e,hunter2006}).

The region NGC6334I contains the well known cometary-shaped
Ultracompact H{\sc ii} (UCH{\sc ii}) region NGC6334F and associated
molecular gas and dust emission. Various maser types were found, from
H$_2$O masers \citep{moran1980,forster1989} and OH masers
\citep{gaume1987,brooks2001} to Class {\sc ii} CH$_3$OH maser emission
\citep{norris1993,caswell1997,walsh1998}. Furthermore, mid-infrared
imaging has identified a potential exciting star for the UCH{\sc ii}
region \citep{debuizer2002}. The main molecular gas and dust peaks are
located at the north-western edge of the UCH{\sc ii} region (e.g.,
\citealt{kraemer1995,beuther2005e,hunter2006}). Multi-frequency
submillimeter (submm) and millimeter (mm) dust continuum imaging
indicates dust temperatures of the order 100\,K \citep{sandell2000}.
Single-dish molecular line surveys revealed a rich line forest (e.g.,
\citealt{mccutcheon2000,thorwirth2003,schilke2006}) comparable to
those observed toward hot core regions like Orion-KL. A molecular
outflow has been observed with a velocity range of approximately
150\,km\,s$^{-1}$ \citep{bachiller1990,leurini2006}.  NH$_3$(3,3)
maser as well as shocked H$_2$ emission was reported toward the end of
the outflow lobes \citep{kraemer1995,davis1995,persi1996,megeath1999}.

The source NGC6334I(N) is located approximately $2'$ to the north of
NGC6334I.  It also exhibits strong (sub)mm continuum emission but at
lower temperatures ($\sim $30\,K). Molecular line surveys of
NGC6334I(N) revealed that most species exhibit fainter emission than
that found toward NGC6334I, although there are some species like
HC$_3$N which are stronger in NGC6334I(N)
\citep{megeath1999,mccutcheon2000,thorwirth2003,sollins2004c}. No
mid-infrared and only weak near-infrared emission has been detected
toward NGC6334I(N) by \citet{tapia1996}. Recent deeper near-infrared
imaging has found evidence for a cluster of low-mass stars in this
region \citep{persi2005}. \citet{carral2002} report the detection of
two faint cm continuum sources, one of them associated with Class {\sc
ii} CH$_3$OH maser emission
\citep{caswell1997,walsh1998}. Furthermore, \citet{kogan1998} observed
a cluster of Class {\sc i} CH$_3$OH masers, and \citet{megeath1999}
detected a bipolar outflow in north-west south-east direction in the
thermal SiO emission. NGC6334I(N) is considered to be the younger of
the two regions.

To better characterize this intriguing pair of massive star-forming
regions, we started an observational campaign with the Australia
Telescope Compact Array (ATCA) and the Submillimeter Array (SMA) from
centimeter (cm) to submm and mm wavelengths.  \citet{beuther2005e}
investigated the regions with the ATCA in the NH$_3$(1,1) and (2,2)
lines as well as CH$_3$OH emission near 25\,GHz.  They found compact
hot gas cores in all lines toward NGC6334I, however, in NGC6334I(N)
only extended NH$_3$ emission, and no thermal CH$_3$OH at the observed
frequencies, was detected.  Toward both regions, the temperatures were
too high to derive reasonable rotational temperatures with the
low-energy-level NH$_3$ transitions.  \citet{hunter2006} identified
multiple mm continuum sources in both regions (four in NGC6334I and
seven in NGC6334I(N)).  Furthermore, an additional north-east
south-west outflow is identified toward NGC6334I(N) (Hunter et al.~in
prep.), oriented approximately perpendicular to the one previously
reported by \citet{megeath1999}.

To get a better understanding of the hot gas components toward the
twin cores NGC6334I and I(N), here we report a follow-up study of the
NH$_3$(3,3) to (6,6) inversion lines with the ATCA.

\section{Observations}

We observed NGC6334I \& I(N) in November 2005 during two nights with
the ATCA in the compact 750D configuration, also including antenna 6.
This results, at 25\,GHz, in projected baselines between 3.8 and
369\,k$\lambda$. The phase reference centers were R.A. (J2000)
$17^h20^m53^s.44$, Decl.~(J2000) $-35^{\circ}47'02''.2$ for NGC6334I
and R.A. (J2000) $17^h20^m54^s.63$, Decl.~(J2000)
$-35^{\circ}45'08''.9$ for NGC6334I(N). We observed the NH$_3$(3,3),
(4,4), (5,5), and (6,6) inversion lines with the frequencies of the
main hyperfine components at 23.870, 24.139, 24.533, and 25.056\,GHz,
respectively. The velocities relative to the local standard of rest
($v_{\rm{lsr}}$) for NGC6334I and NGC6334I(N) are $\sim -7.6$ and
$\sim -3.3$\,km\,s$^{-1}$, respectively. Good uv-coverage was
obtained through regular switching between both sources and the four
spectral setups. On-source integration times for each of the
NH$_3$(3,3) and (4,4) lines, in both sources, were 220 minutes.  The
(5,5) and (6,6) lines were observed each for 200 min in NGC6334I and
for 190 min in NGC6334I(N). The spectral resolution of the
observations was 62\,kHz, corresponding to a velocity resolution of
$\sim 0.8$\,km\,s$^{-1}$.  The primary beam of the ATCA at the given
frequency is $\sim 130''$.  The data were reduced with the MIRIAD
package.  Applying a robust weighting of 1 (closer to natural than
uniform weighting, thus stressing the shorter baselines) the
synthesized beam is $1.9''\times 1.1''$. The rms per 1\,km\,s$^{-1}$
channel is $\sim$3\,mJy.

\section{Results}

\begin{figure*}[htb]
\includegraphics[angle=-90,width=8.8cm]{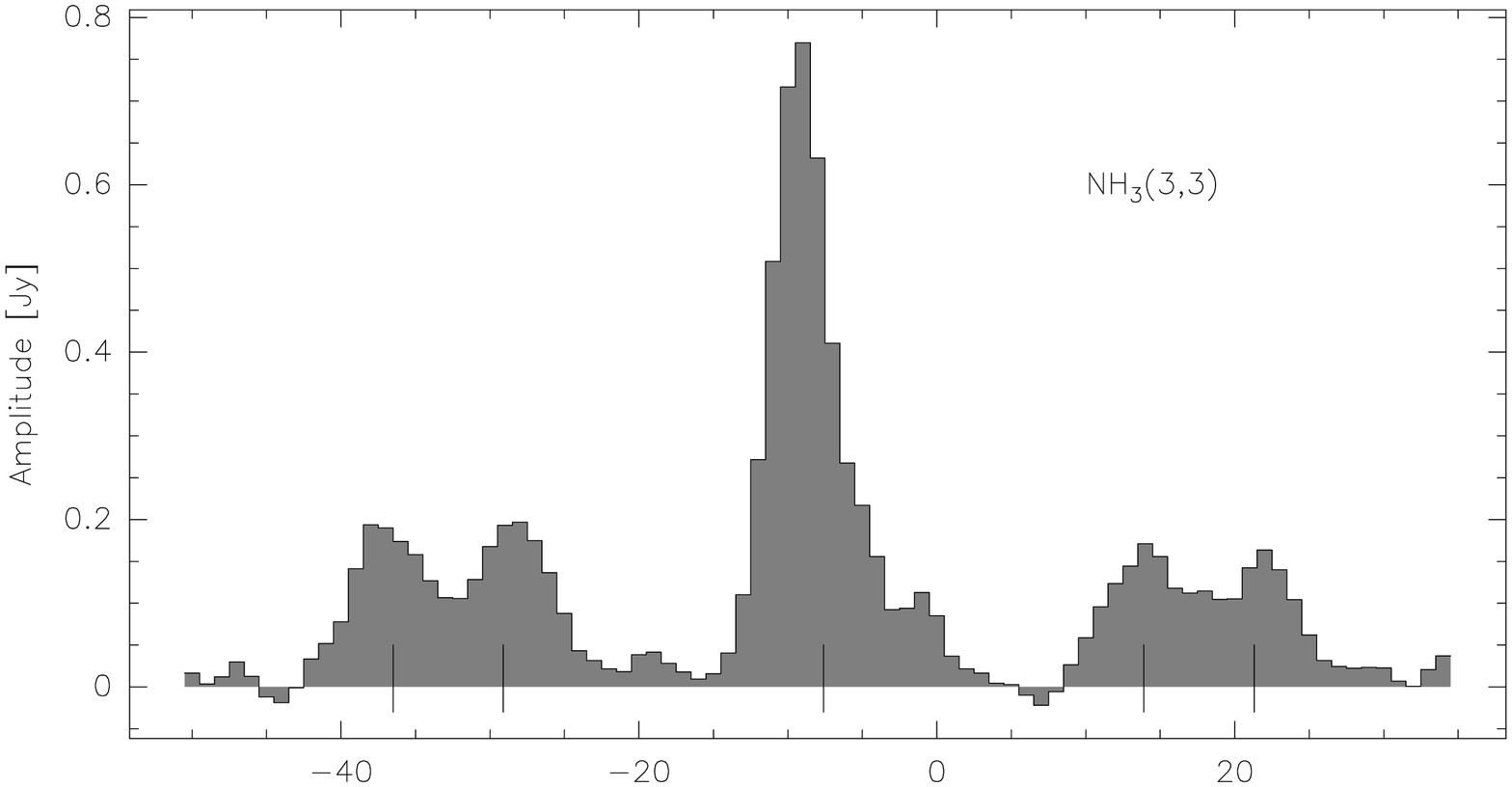}
\includegraphics[angle=-90,width=8.8cm]{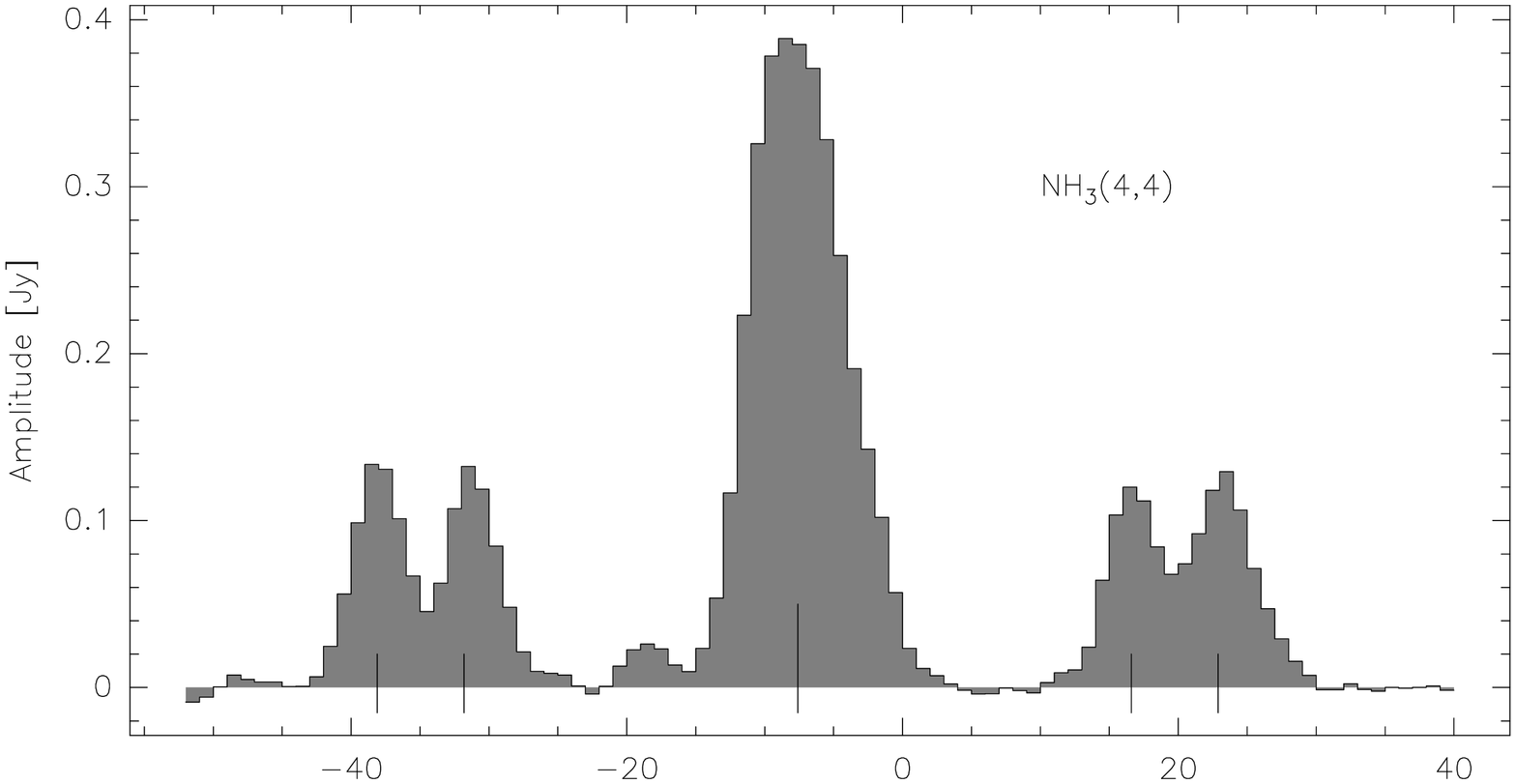}\\
\includegraphics[angle=-90,width=8.8cm]{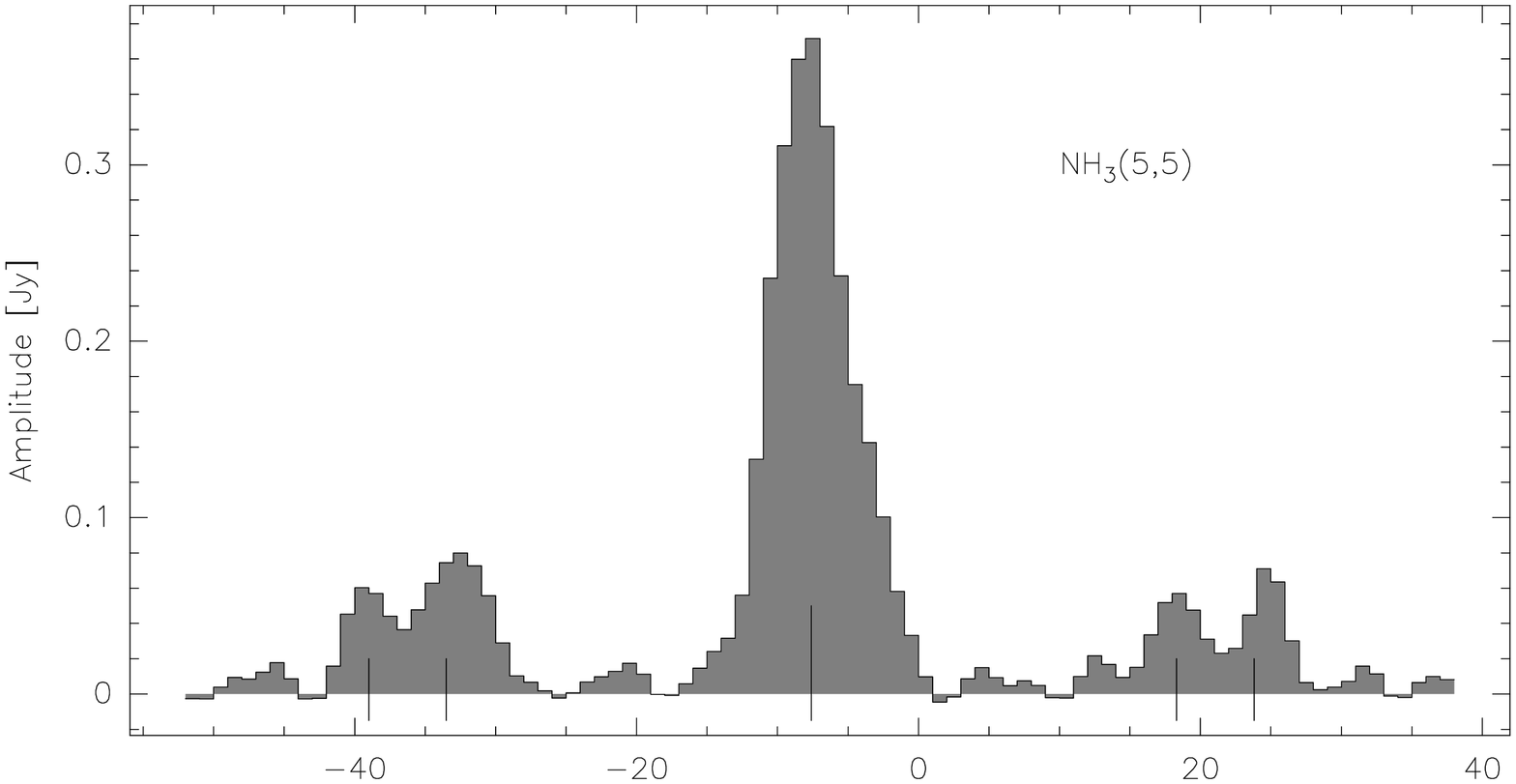}
\includegraphics[angle=-90,width=8.8cm]{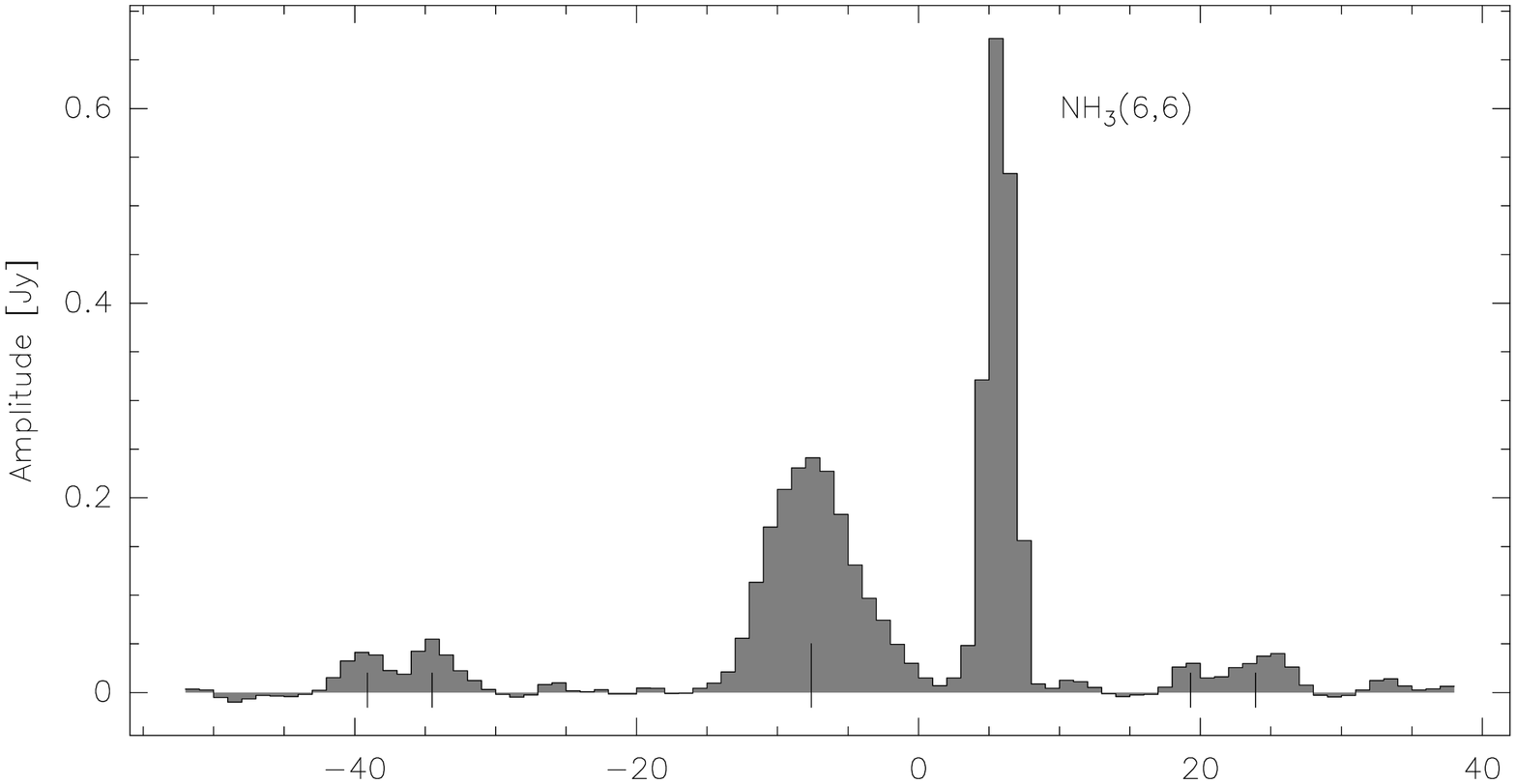}
\caption{UV-Spectra of the NH$_3$ inversion lines on the shortest
  baseline toward NGC6334I. The lines mark the spectral positions of the main and satellite hyperfine components.}
\label{ngc6334i_uvspectra}
\end{figure*}

\begin{figure*}[htb]
\includegraphics[angle=-90,width=8.8cm]{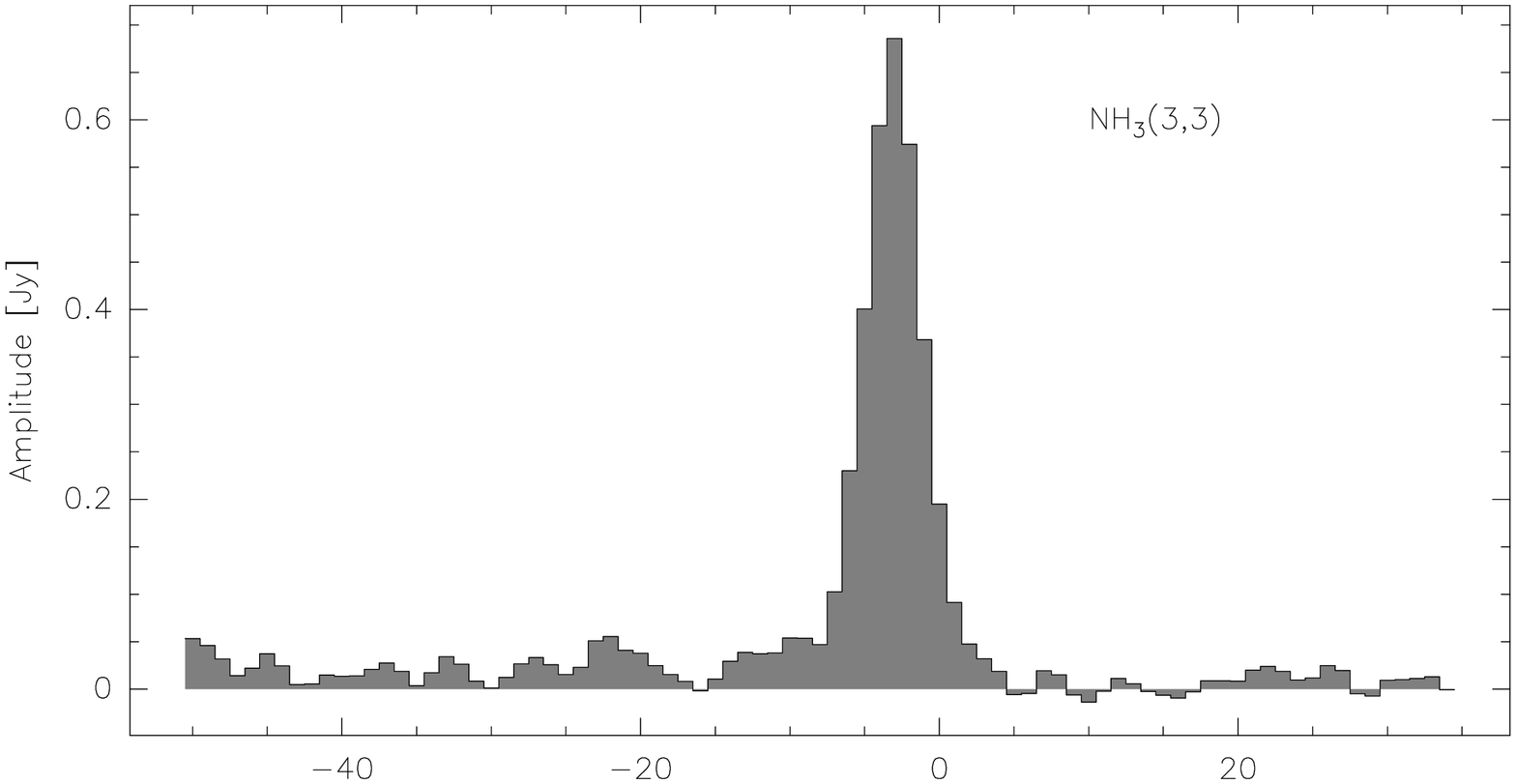}
\includegraphics[angle=-90,width=8.8cm]{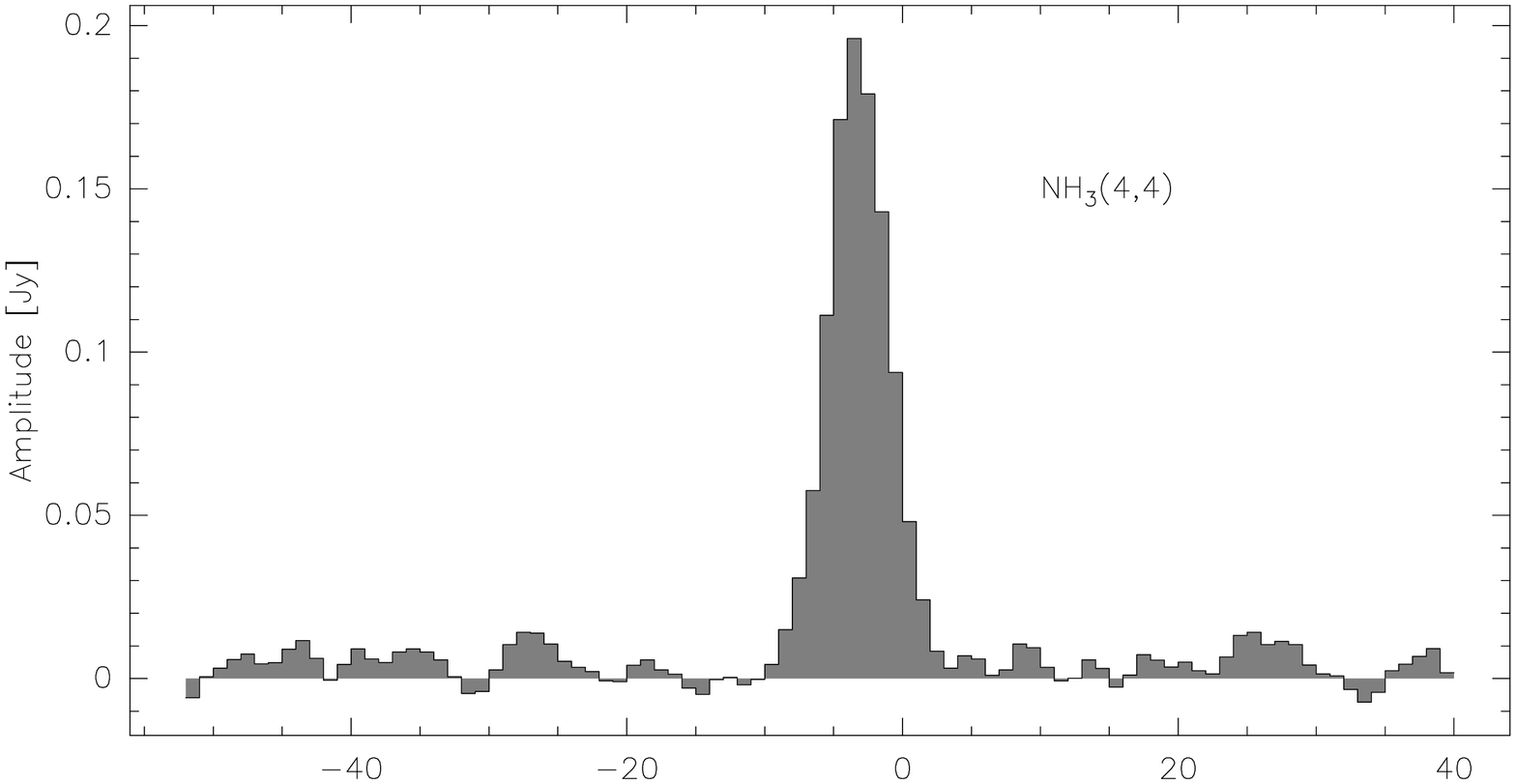}\\
\includegraphics[angle=-90,width=8.8cm]{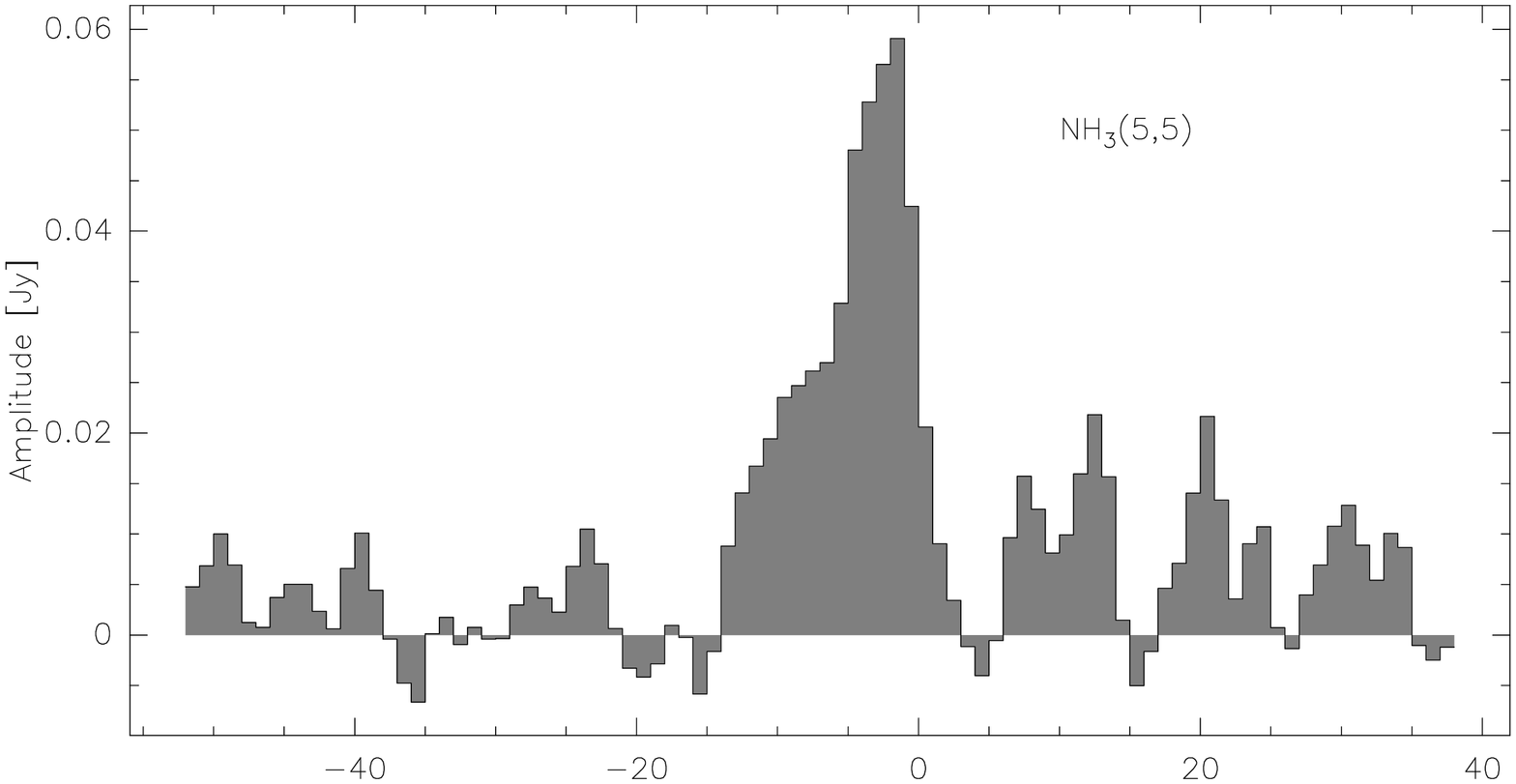}
\includegraphics[angle=-90,width=8.8cm]{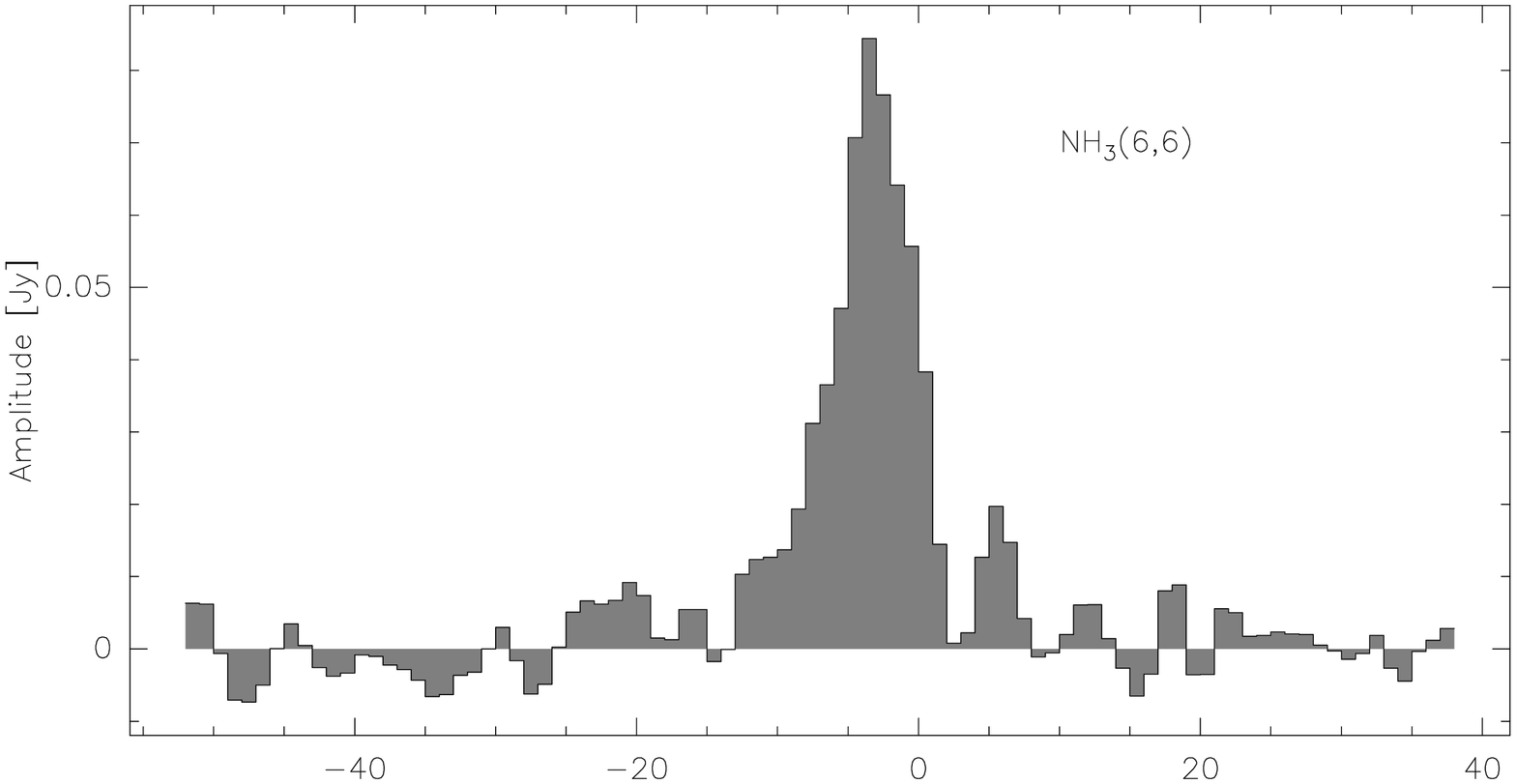}
\caption{UV-Spectra of the NH$_3$ inversion lines on the shortest
  baseline toward NGC6334I(N).}
\label{ngc6334in_uvspectra}
\end{figure*}

The four observed NH$_3$ inversion lines cover a broad range of energy
levels above ground with lower energy levels $E_l$ between 123 and
407\,K ($E_l(\rm{NH_3(3,3)}=123$\,K, $E_l(\rm{NH_3(4,4)}=200$\,K,
$E_l(\rm{NH_3(5,5)}=295$\,K, $E_l(\rm{NH_3(6,6)}=407$\,K). The
previous NH$_3$(1,1) and (2,2) data of \citet{beuther2005e} extend the
range of observed energy levels down to 23\,K.  Hence, we are able to
trace various temperature components throughout the two star-forming
regions. All 6 inversion lines up to NH$_3$(6,6) were detected in both
target regions (Figs.~\ref{ngc6334i_uvspectra} \&
\ref{ngc6334in_uvspectra}).  While this is less surprising for
NGC6334I, the northern source NGC6334I(N) lacks strong infrared
emission and is considered a prototypical massive cold core at a very
early stage of star formation (e.g., \citealt{gezari1982}). While it
is still considered to be a very young source, the detection of the
high energy level transitions up to NH$_3$(6,6) clearly shows that
compact warm gas components already exist at this early evolutionary
stage. We discuss both regions separately first.

\subsection{NGC6334I}

\subsubsection{Hot thermal NH$_3$ emission}
\label{thermal}

\begin{figure*}[htb]
\includegraphics[angle=-90,width=17.8cm]{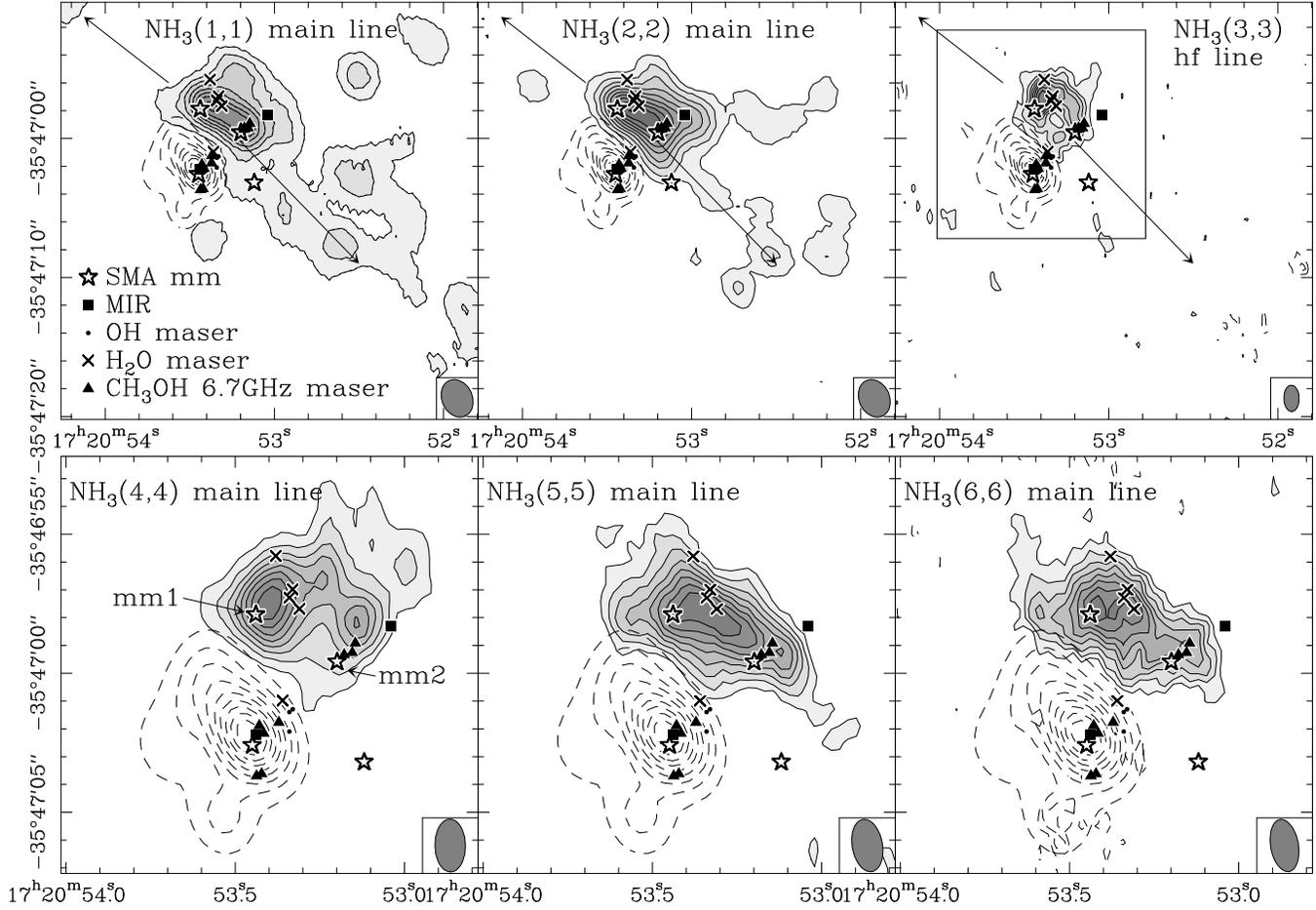}
\caption{Integrated NH$_3$ inversion line images of the thermal
  emission toward NGC6334I are shown in grey-scale with solid contours.
  The (1,1) and (2,2) images are adapted from \citet{beuther2005e}.
  In all cases we show the integrated main line (see
  Fig.~\ref{ngc6334i_uvspectra}), except for the (3,3) transition
  where we present the emission integrated over the two hyperfine
  satellite components at blue-shifted velocities (see
  Fig.\ref{ngc6334i_uvspectra}). The top-row images are shown on
  larger spatial scales than the bottom-row images, the inlay size for
  the bottom-row images is shown in the (3,3) panel.  The contouring
  of the (1,1) and (2,2) is done in $3\sigma$ steps with $3\sigma$
  values of 21 and 12\,mJy\,beam$^{-1}$, respectively.  The (3,3) to
  (6,6) integrated images are contoured from 20 to 90\% (step 10\%) of
  the peak emission with peak values of 22.5, 32.6, 24.2 and
  17.5\,mJy\,beam$^{-1}$, respectively. The dashed contours show the
  Ultracompact H{\sc ii} region from \citet{beuther2005e}, contoured
  from 10 to 90\% (step 10\%) of the peak emission of
  1197\,mJy\,beam$^{-1}$. The markers are labeled in the top row, SMA
  mm continuum emission is from \citet{hunter2006}, H$_2$O maser
  emission from \citet{forster1989}, CH$_3$OH class {\sc ii} maser
  emission from \citet{walsh1998}, OH maser emission from
  \citet{brooks2001} and MIR sources from \citet{debuizer2002}. The
  two main mm continuum sources mm1 and mm2 are labeled in the
  bottom-left panel. The arrows in the top row mark the direction of
  the CO/SiO outflow.  The synthesized beams are shown at the
  bottom-right of each panel. {\it The data to this Figure are also
    available in electronic form at the CDS via anonymous ftp to
    cdsarc.u-strasbg.fr (130.79.128.5) or via
    http://cdsweb.u-strasbg.fr/cgi-bin/qcat?J/A+A/.}}
\label{ngc6334i_images}
\end{figure*}

The uv-spectra taken on the shortest baseline in NGC6334I
(Fig.\ref{ngc6334i_uvspectra}) are rare examples showing even the
satellite hyperfine components, in the highly excited NH$_3$(5,5) and
(6,6) lines, in emission ($\sim$26 and $\sim$31\,km\,s$^{-1}$ offset
from the $v_{\rm{lsr}}$).  Figure \ref{ngc6334i_images} presents
integrated images of all 6 NH$_3$ inversion transitions from the (1,1)
to the (6,6) transition (the (1,1), (2,2) and cm continuum images are
adapted from \citealt{beuther2005e}). Except for the (3,3) line, we
show the integrated emission of the main central hyperfine line. Only
for the (3,3) transition we integrated over the two hyperfine
components at higher frequencies (negative velocities) because the
main line is contaminated by maser emission (see \citealt{kraemer1995}
and section \ref{maser} below).

The NH$_3$(1,1) and (2,2) maps show two peaks associated with the two
main mm continuum sources, mm1 and mm2, reported by
\citet{hunter2006}.  While the two peaks have approximately the same
intensities in the low-energy NH$_3$ lines, the mm dust continuum flux
density of mm1 is approximately 1.5 times higher than that of mm2
\citep{hunter2006}.  The NH$_3$(3,3) hyperfine map shows emission
toward the north-eastern mm source, mm1, and only weaker emission
toward the south-western mm source, mm2. Re-inspecting the NH$_3$(1,1)
and (2,2) data, integrated maps of their satellite hyperfine lines are
also stronger toward mm1 than toward mm2, although the main hyperfine
lines show the same intensities toward both mm peaks. This indicates
that the NH$_3$ optical depth toward mm2 has to be lower than toward
mm1. The higher NH$_3$ inversion lines all show a bright emission peak
near mm1, and a much weaker peak near mm2. Particularly in the
NH$_3$(4,4) and (5,5) lines, this weaker peak is shifted to the west
of mm2 toward the mid-infrared source detected by
\citet{debuizer2002}. Although in the lower energy NH$_3$ lines the
two mm peaks appear similar, the fact that the higher energy lines are
much stronger toward mm1 indicates that this is the warmer of two
sources.

\begin{figure*}[htb]
\includegraphics[angle=-90,width=8.8cm]{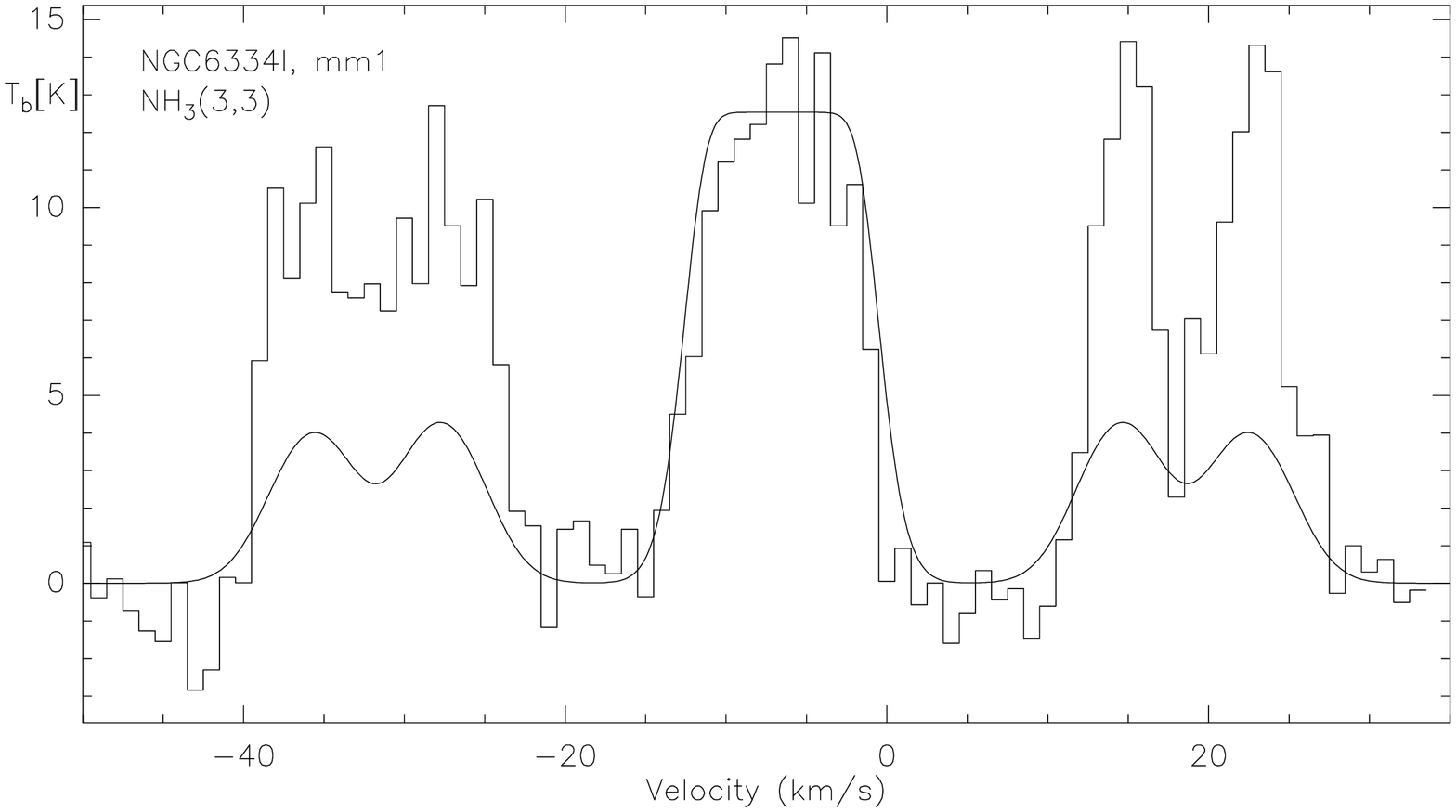}
\includegraphics[angle=-90,width=8.8cm]{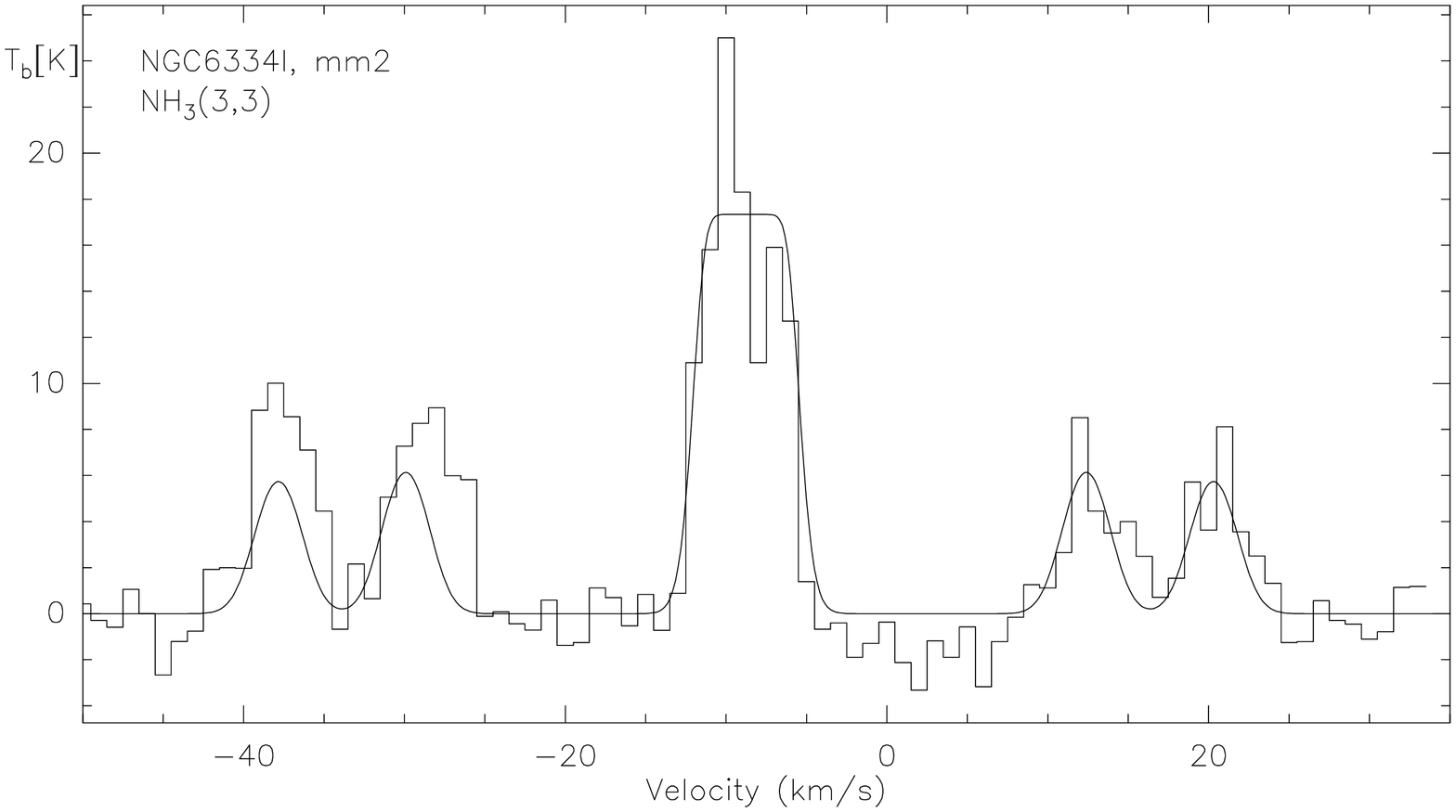}\\
\includegraphics[angle=-90,width=8.8cm]{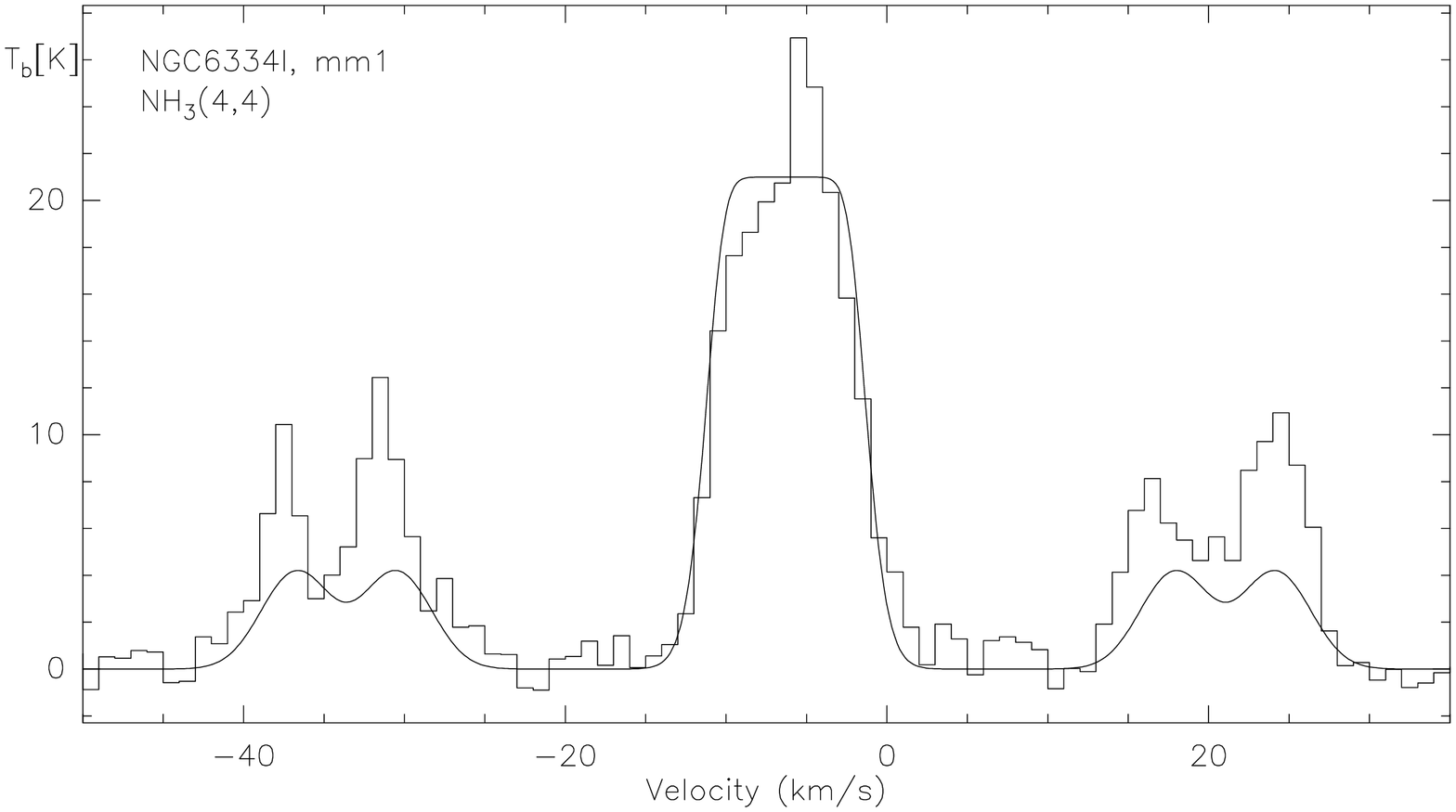}
\includegraphics[angle=-90,width=8.8cm]{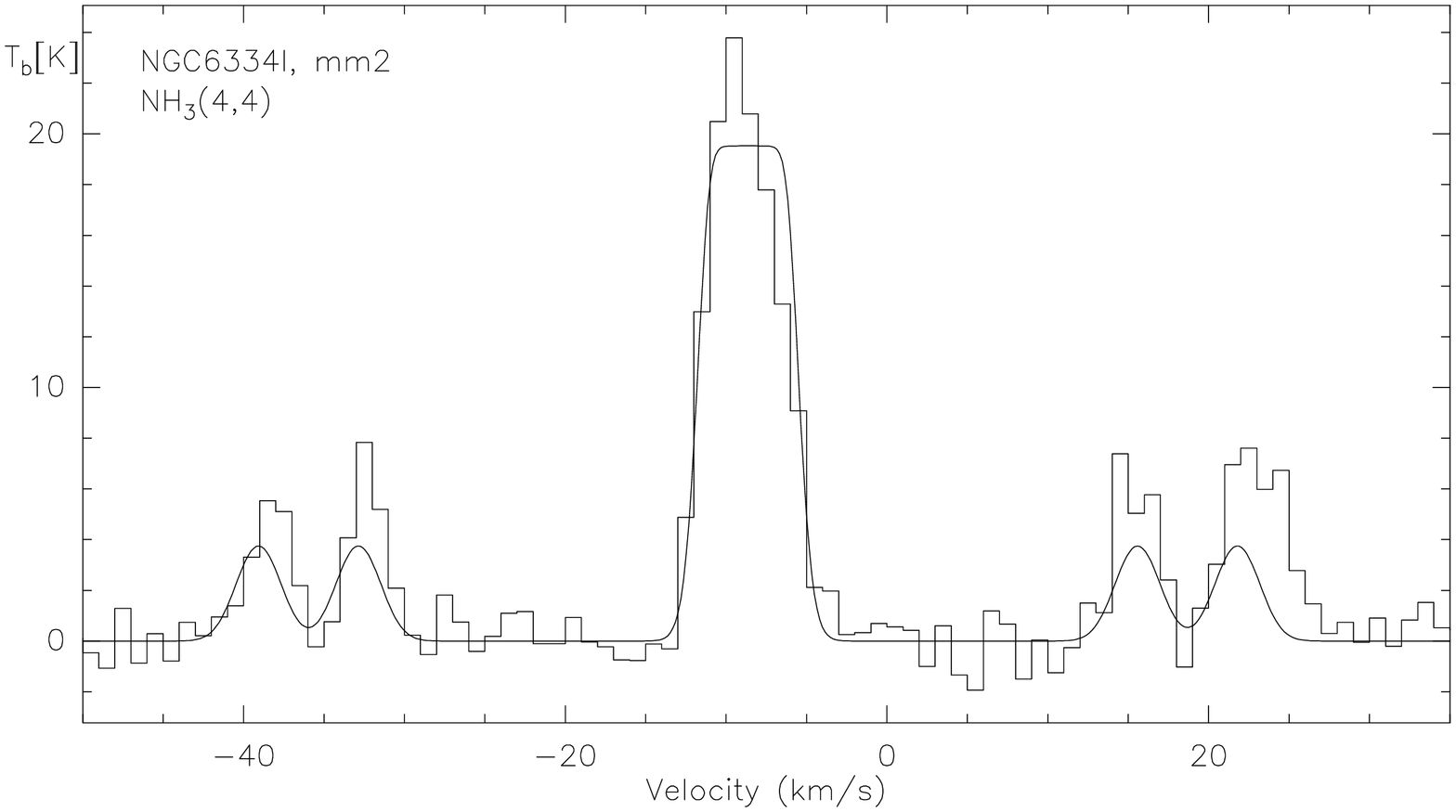}\\
\includegraphics[angle=-90,width=8.8cm]{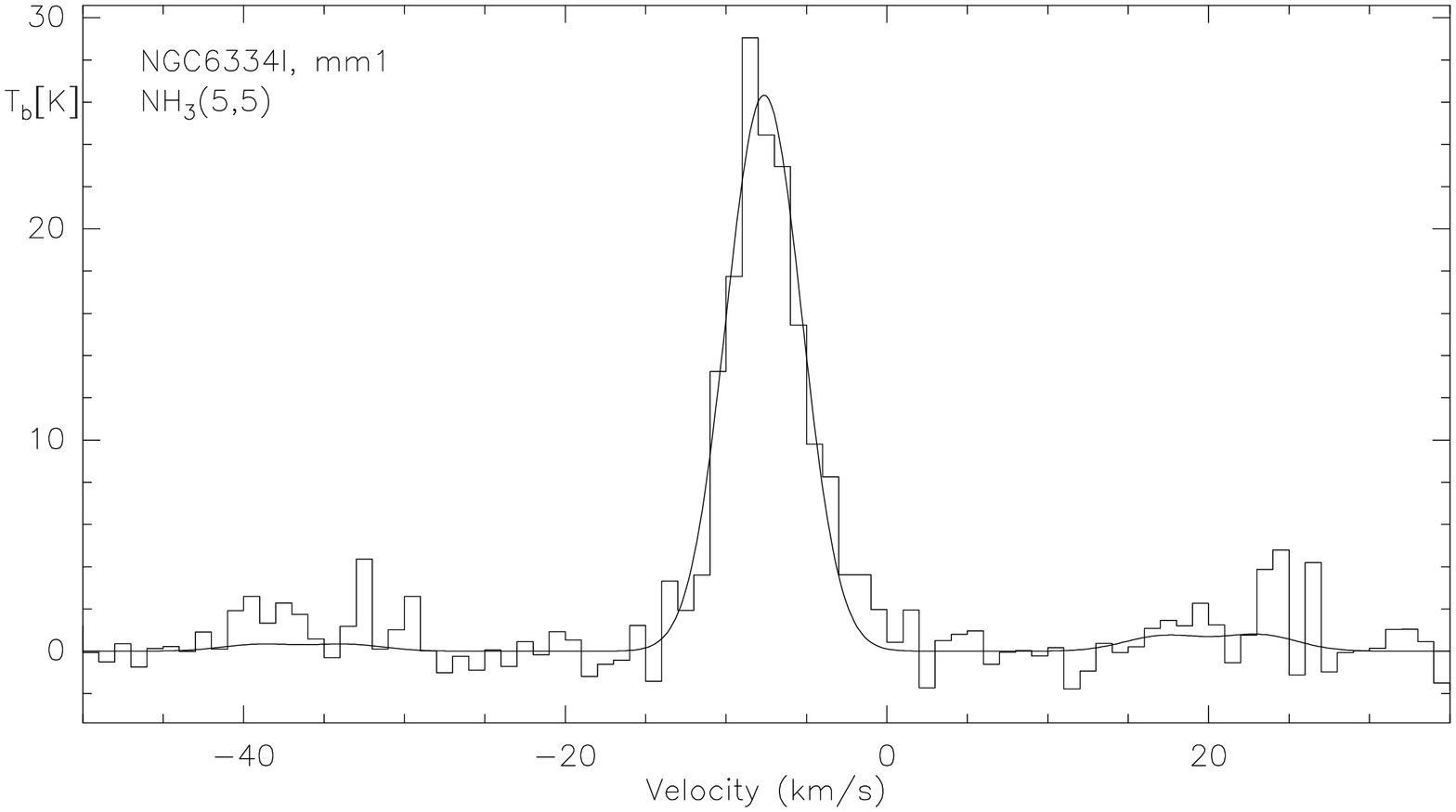}
\includegraphics[angle=-90,width=8.8cm]{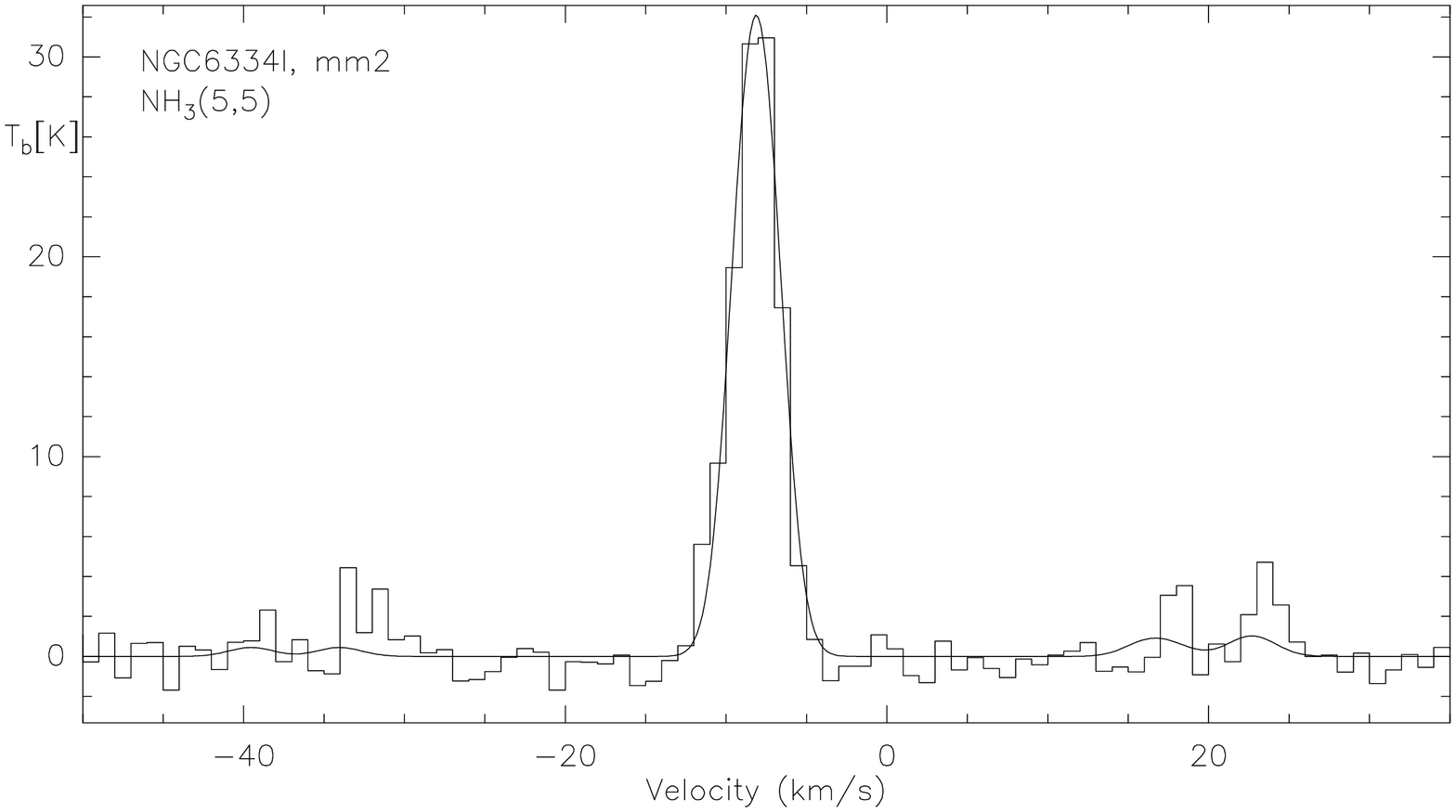}\\
\includegraphics[angle=-90,width=8.8cm]{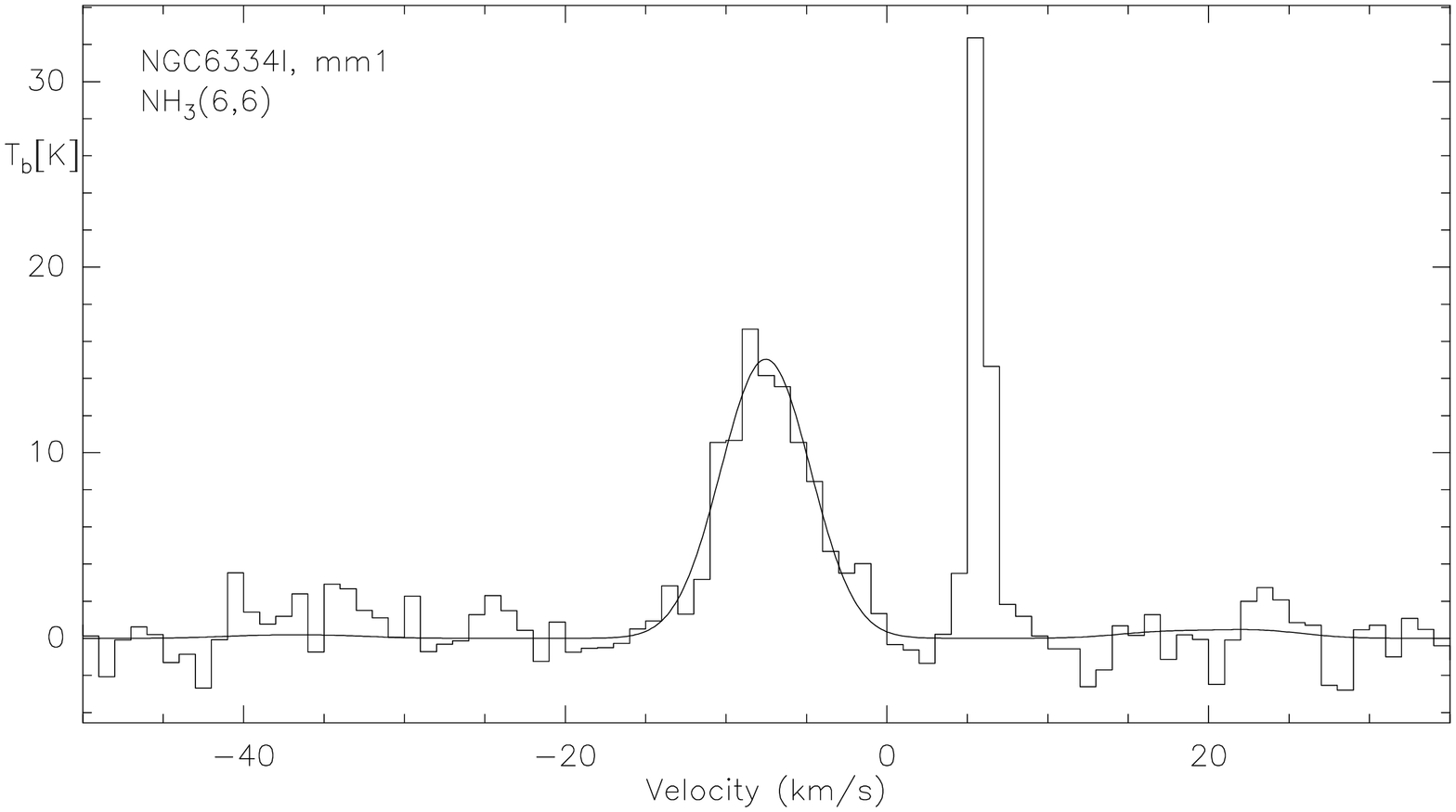}
\includegraphics[angle=-90,width=8.8cm]{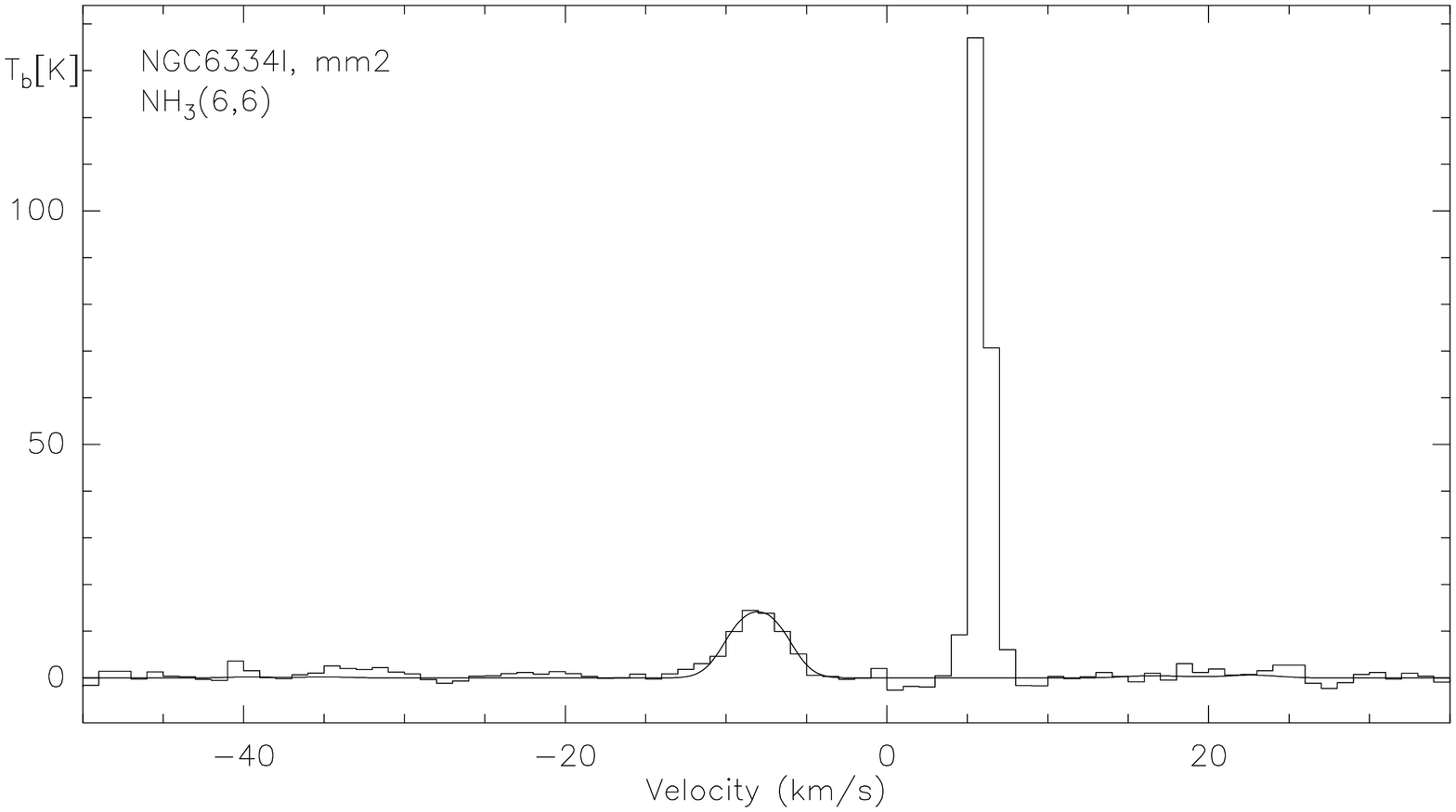}
\caption{NH$_3$(3,3) to (6,6) spectra toward the molecular peak
  positions associated with the mm continuum peaks mm1 and mm2. The
  spectral resolution is smoothed to 1\,km\,s$^{-1}$ and the spatial
  resolution to $2.8''\times 2.2''$ (the same as the spatial
  resolution of the lower energy lines presented in
  \citealt{beuther2005e}). The additional strong peaks at positive
  velocities in the NH$_3$(6,6) spectra are a maser discussed in \S
  \ref{maser}. The solid lines show fits to the spectra. The (3,3) and
  (4,4) lines are optically thick so that reasonable fits are not
  possible. {\it The data to this Figure are also available in
    electronic form at the CDS via anonymous ftp to
    cdsarc.u-strasbg.fr (130.79.128.5) or via
    http://cdsweb.u-strasbg.fr/cgi-bin/qcat?J/A+A/.}}
\label{spectra_i}
\end{figure*}

One of the aims of this study is to get temperature estimates of the
warm gas. We identify the positions of the two main molecular line
peaks with the NH$_3$(6,6) emission peak (R.A. (J2000) 17:20:53.42,
Dec (J2000) -35:46:57.7, offset $-0.3''/4.5''$ from the phase center)
near mm1, and one Class {\sc ii} CH$_3$OH maser feature (R.A. (J2000)
17:20:53.18, Dec (J2000) -35:46:59.3, \citealt{caswell1997,walsh1998},
offset $-3.2''/2.9''$ from the phase center) near mm2. The spectra of
all four newly observed lines toward these two positions are presented
in Fig.~\ref{spectra_i}. Fitting of the hyperfine structure was done
in CLASS which is part of the GILDAS software suite\footnote{See
  http://www.iram.fr/IRAMFR/GILDAS}. Recent frequency measurements for
NH$_3$ and its deuterated variants \citep{coudert2006} have been taken
into account.

Toward both positions, the NH$_3$(3,3) and (4,4) lines are optically
thick so we are not able to derive reasonable fits with the standard
fitting procedures assuming constant excitation temperatures. This way
we get only flat-topped synthetic spectra.  Since such flat-topped
spectra are not observed, this indicates that different temperatures
likely exist along the line of sight. The most extreme case is the
NH$_3$(3,3) spectrum toward NGC6334I-mm1 (Fig.~\ref{spectra_i}
top-left panel): In the optically thin case the peak intensities of
the satellite hyperfine lines should be 3\% of the peak intensity of
the main line. However, in this case the satellite hyperfine lines of
the NH$_3$(3,3) transition have more or less the same intensities as
the main line, indicating the large optical depth.  Since we cannot
fit these lines well, we use the higher excitation NH$_3$(5,5) and
(6,6) lines for the temperature estimates.

Reasonable fits are possible for these two lines, and we measure the
peak intensities, the opacities of the main lines, and the line-widths
(Table \ref{linefits}).  The spectral peak intensities toward these
two positions are similar, which is in contrast to the integrated
images shown in Figure \ref{ngc6334i_images}. However, the line-widths
toward mm1 are more than 2\,km\,s$^{-1}$ larger than those toward mm2;
this accounts for the differences in the integrated images.  Because
the NH$_3$(5,5) and (6,6) lines are of different NH$_3$ species
(para-NH$_3$ and ortho-NH$_3$, respectively), their different
statistical weights have to be taken into account.  Doing this, we can
derive rotational temperatures ($T_{\rm{rot}}$) toward the molecular
peak positions associated with mm1 and mm2. The derived
$T_{\rm{rot}}(55,66)$ values are 86$\pm 20$\,K and 67$\pm 20$\,K for
the two positions, respectively. Although the temperatures are
comparable within the error-bars, the data are indicative that
temperatures associated with mm1 may be higher than those associated
with mm2.

\begin{table}[htb]
\caption{Fitted parameters and rotational temperatures in NGC6334I.}
\begin{center}
\begin{tabular}{lrr}
\hline \hline
      & mm1 & mm2 \\
\hline
Offset [$''$] & -0.3/4.5 & -3.2/2.9 \\
$T_{\rm{b}}$(NH$_3$(5,5)) [K] & 26.4$\pm 0.3$ & 32.2$\pm 0.3$\\
$\Delta v$(NH$_3$(5,5)) [km\,s$^{-1}$] & 5.4$\pm 0.2$ & 3.3$\pm 0.2$ \\
$v_{\rm{peak}}$(NH$_3$(5,5)) [km\,s$^{-1}$] & -7.6$\pm 0.1$ & -8.1$\pm 0.1$ \\
$\tau_{\rm{main}}$(NH$_3$(5,5))$^a$ & 0.1$\pm 1.8$ & 0.4$\pm 0.5$ \\
$T_{\rm{b}}$(NH$_3$(6,6)) [K] & 15.1$\pm 0.3$ & 14.3$\pm 0.3$ \\
$\Delta v$(NH$_3$(6,6)) [km\,s$^{-1}$] & 6.4$\pm 0.6$ & 3.5$\pm 0.6$ \\
$v_{\rm{peak}}$(NH$_3$(6,6)) [km\,s$^{-1}$] & -7.5$\pm 0.2$ & -8.0$\pm 0.1$ \\
$\tau_{\rm{main}}$(NH$_3$(6,6))$^a$ &  0.1$\pm 2.8$  & 1.4$\pm 1.6$ \\
$T_{\rm{rot}}(5,5/6,6)$$^b$ [K] & 86$\pm 20$ & 67$\pm 20$ \\ 
\hline \hline
\end{tabular}
\end{center}
$^a$ $\tau_{\rm{main}}$ is prone to the largest errors, since the fits 
in Fig.~\ref{spectra_i} do not accurately reproduce the satellite lines.\\
$^b$ The  errors for $T_{\rm{rot}}$ are estimated by varying 
$\tau_{\rm{main}}$ within errors.
\label{linefits}
\end{table}

For the very optically thick lines where the previous fits failed, it
is also possible to estimate approximate opacities $\tau_{\rm{main}}$
from the observed line ratios between the main and the satellite
hyperfine components. Following \citet{ho1983}, we then calculated the
rotational temperatures, $T_{\rm{rot}}$, and NH$_3$ column
densities. Table \ref{thick} list the values derived via this
approach. The estimated opacities of the (3,3) and (4,4) lines are all
significantly larger than those we derived from the (5,5) and (6,6)
lines. Interestingly, the rotational temperature for mm2 corresponds
well to the rotational temperature derived from the higher excited
line pair above.  In contrast to that, the temperature we now derived
for mm1 is higher than that derived from the higher excitation lines
by a factor of about 1.6.  While the estimated errors associated with
the the (3,3)/(4,4) line pair are even larger than for the (5,5)/(6,6)
pair, this derivation confirms the previous assessment that mm1 is
likely the warmer of the two sources. The NH$_3$ column densities
derived this way would correspond to H$_2$ column densities of the
order a few times $10^{26}$\,cm$^{-2}$ (assuming NH$_3$ Orion hot core
abundances as listed in \citealt{vandishoek1998}). This is about 1-2
orders of magnitude larger than the H$_2$ column densities one can
derive at comparable spatial scales from the mm continuum data
\citep{hunter2006}. This is an additional indicator of the limitations
one has to face when the lines have as high opacities as in the case
of NGC6334I. Future radiative transfer modeling may explain these
observed spectral lines better; however, this is beyond the scope of
this paper.

\begin{table}[htb]
\caption{Parameters for NGC6334I derived via the satellite/main line ratios.}
\begin{center}
\begin{tabular}{lrr}
\hline \hline
      & mm1 & mm2 \\
\hline
Offset [$''$] & -0.3/4.5 & -3.2/2.9 \\
$\tau_{\rm{main}}$(NH$_3$(3,3)) & 64$\pm 5$ & 13$\pm 5$ \\
$\tau_{\rm{main}}$(NH$_3$(4,4)) & 28$\pm 5$ & 20$\pm 5$ \\
$T_{\rm{rot}}(3,3/4,4)$ [K] & 140$\pm 50$ & 70$\pm 50$ \\ 
$N(\rm{NH_3})$$^a$ [cm$^{-2}$] & $1.0\times 10^{19}\pm 30\%$ & $6.8\times 10^{18}\pm 30\%$\\ 
\hline \hline
\end{tabular}
\end{center}
\footnotesize{$^a$ Since we cannot derive reasonable line-widths for these lines, for the column density calculations we assume $\Delta v=5.2$ and $\Delta v=3.2$ for mm1 and mm2, respectively (see Table \ref{linefits}).}
\label{thick}
\end{table}

The NH$_3$(6,6) line-width is larger toward mm1 than the
NH$_3$(5,5) line-width. For mm2, they are approximately the same
within the errorbars. It would be interesting to compare these
line-widths also with those of the lower excitation lines.  Since, we
cannot accurately fit the whole spectra (see Fig.~\ref{spectra_i}), we
tried to derive the line widths via Gaussian fits to the satellite
hyperfine components. However, for the same transitions, Gaussian fits
of various satellite hyperfine components result in different
line-widths (differences up to 1\,km\,s$^{-1}$) indicative of the high
optical depth and the complex source structure that is also evident in
the various hyperfine components.  Therefore, a comparison of the
line-widths of the lower-excitation lines is difficult.  Nevertheless,
toward mm1 the larger line-width of the higher excitation NH$_3$(6,6)
line compared to NH$_3$(5,5) indicates larger internal motions of the
warmer gas which is likely closer to the central source.

\subsubsection{NH$_3$(3,3) and (6,6) maser emission}
\label{maser}

\begin{figure*}[htb]
\includegraphics[angle=-90,width=17.8cm]{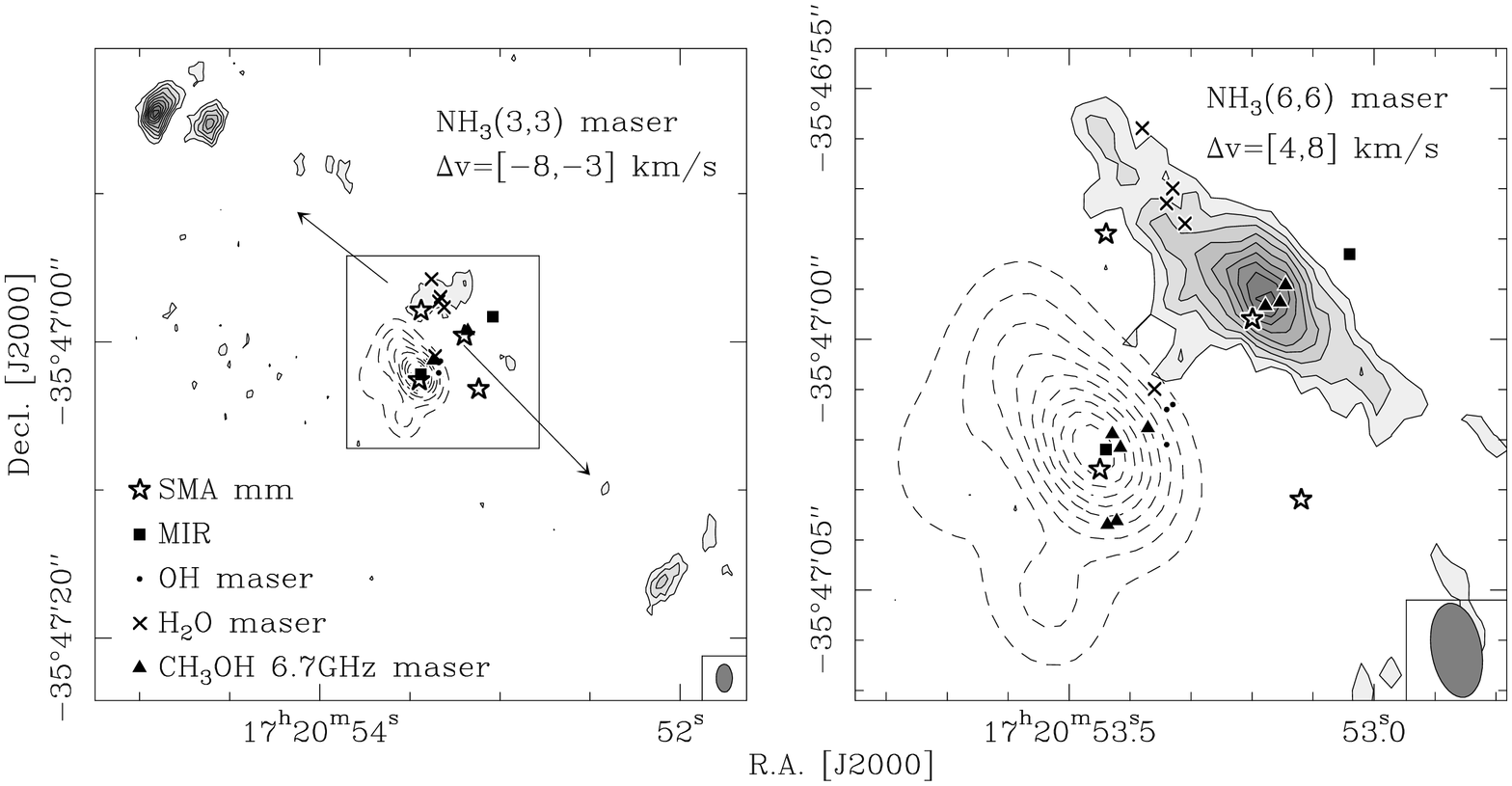}
\caption{NH$_3$(3,3) and (6,6) maser emission in NGC6334I is shown in
  the left and right panel, respectively. Velocity ranges of
  integrated emission are given in each panel and the markers are the
  same as in Fig.~\ref{ngc6334i_images}.  The contour levels are from
  15 to 95\% (step 10\%) of the peak emission, which is 112.5 and
  126.8\,mJy\,beam$^{-1}$ for the (3,3) and (6,6) maser lines.  The
  dashed contours show the Ultracompact H{\sc ii} region from
  \citet{beuther2005e}, contoured from 10 to 90\% (step 10\%) of the
  peak emission of 1197\,mJy\,beam$^{-1}$. The right panel again shows
  a smaller region as shown by the inlay in the left panel. The
  synthesized beams are shown at the bottom-right of each panel. {\it
    The data to this Figure are also available in electronic form at
    the CDS via anonymous ftp to cdsarc.u-strasbg.fr (130.79.128.5) or
    via http://cdsweb.u-strasbg.fr/cgi-bin/qcat?J/A+A/.}}
\label{ngc6334i_maser}
\end{figure*}

\begin{figure}[htb]
\includegraphics[angle=-90,width=8cm]{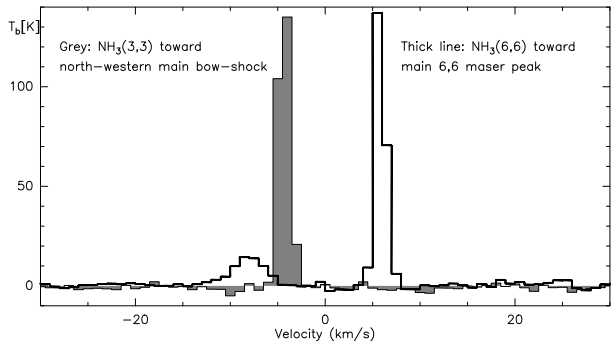}
\caption{NH$_3$(3,3) and (6,6) maser spectra in grey and thick lines,
  respectively. The (3,3) spectrum is extracted toward the main
  north-eastern bow-shock, and the (6,6) spectra is taken toward the
  main (6,6) maser peak close to mm2. (see Fig.~\ref{ngc6334i_maser}).
  {\it The data to this Figure are also available in electronic form
    at the CDS via anonymous ftp to cdsarc.u-strasbg.fr (130.79.128.5)
    or via http://cdsweb.u-strasbg.fr/cgi-bin/qcat?J/A+A/.}}
\label{maserspectra}
\end{figure}

\citet{kraemer1995} reported NH$_3$(3,3) maser emission toward the
outflow lobes in NGC6334I. This observation is confirmed by our data
(Fig.~\ref{ngc6334i_maser} \& \ref{maserspectra}). The most
north-eastern (3,3) maser spot is resolved and shows a morphology 
expected from bow-shocks in molecular outflows. A Gaussian fit to the
spectrum extracted toward the bow-shocks results in a peak-brightness
temperature $T_b\sim 160$\,K and a linewidth $\Delta v\sim
1.6$\,km\,s$^{-1}$. The maser emission is identified by a combination
of high brightness temperatures, narrow line-widths and peculiar
spatial distributions, all with respect to the known thermal gas
components. This agrees with the argument set forth by
\citet{kraemer1995} that the NH$_3$(3,3) maser emission is not
associated with any other maser types, but that it is likely caused by
shocks when the outflow impinges on the ambient molecular gas.

Here, we report NH$_3$(6,6) maser emission toward NGC6334I
(Figs.~\ref{spectra_i}, \ref{ngc6334i_maser} \& \ref{maserspectra}).
We believe that this is the first detection of NH$_3$(6,6) maser
emission. The NH$_3$(6,6) maser is not associated with the NH$_3$(3,3)
maser emission toward the outflow lobes, but rather, the (6,6) maser
emission is found at the core center and peaks coincident with the
Class {\sc ii} CH$_3$OH maser group associated with mm2.  The maser
emission is resolved and elongated in the north-east south-western
direction approximately along the large-scale outflow axis. This
apparent extended structure is likely a superposition of several
point-like sub-features which cannot be resolved by our observations.
Weak NH$_3$(6,6) maser emission is found in the vicinity of mm1 and
associated with the H$_2$O maser cluster.  The $v_{\rm{lsr}}$ of the
(6,6) maser is +5.8\,km\,s$^{-1}$ approximately 13\,km\,s$^{-1}$
offset from the thermal emission at that position.  The FWHM of the
maser feature is $\sim 1.3$\,km/s, far smaller than the line-width of
the thermally excited lines reported in Table~\ref{linefits}.  The
H$_2$O maser velocities reported by \citep{forster1989} range from
$-45.6$ to $-0.8$\,km\,s$^{-1}$ and are hence also offset from the
NH$_3$(6,6) maser velocity. Although the Class {\sc ii} CH$_3$OH maser
peaks are at the same positions, their velocities (from $-6$ to
$-12$\,km\,s$^{-1}$, \citealt{walsh1998}) are also offset from the
NH$_3$(6,6) maser velocity.  Unfortunately, from a theoretical point
of view, the NH$_3$(6,6) maser has not been studied yet, and we thus
have no predictions of what to expect. As discussed in
\S\ref{outflowdriving}, the NH$_3$(6,6) maser morphology indicates
that its emission is associated with the molecular outflow. Although
we do not yet understand the peculiar velocity of the (6,6) maser,
this indicates that it is likely associated with some outflow-shock
processes.  Further theoretical and observational work is required to
understand its characteristics in better detail.

\subsection{NGC6334I(N)}

\begin{figure*}[htb]
\includegraphics[angle=-90,width=17.8cm]{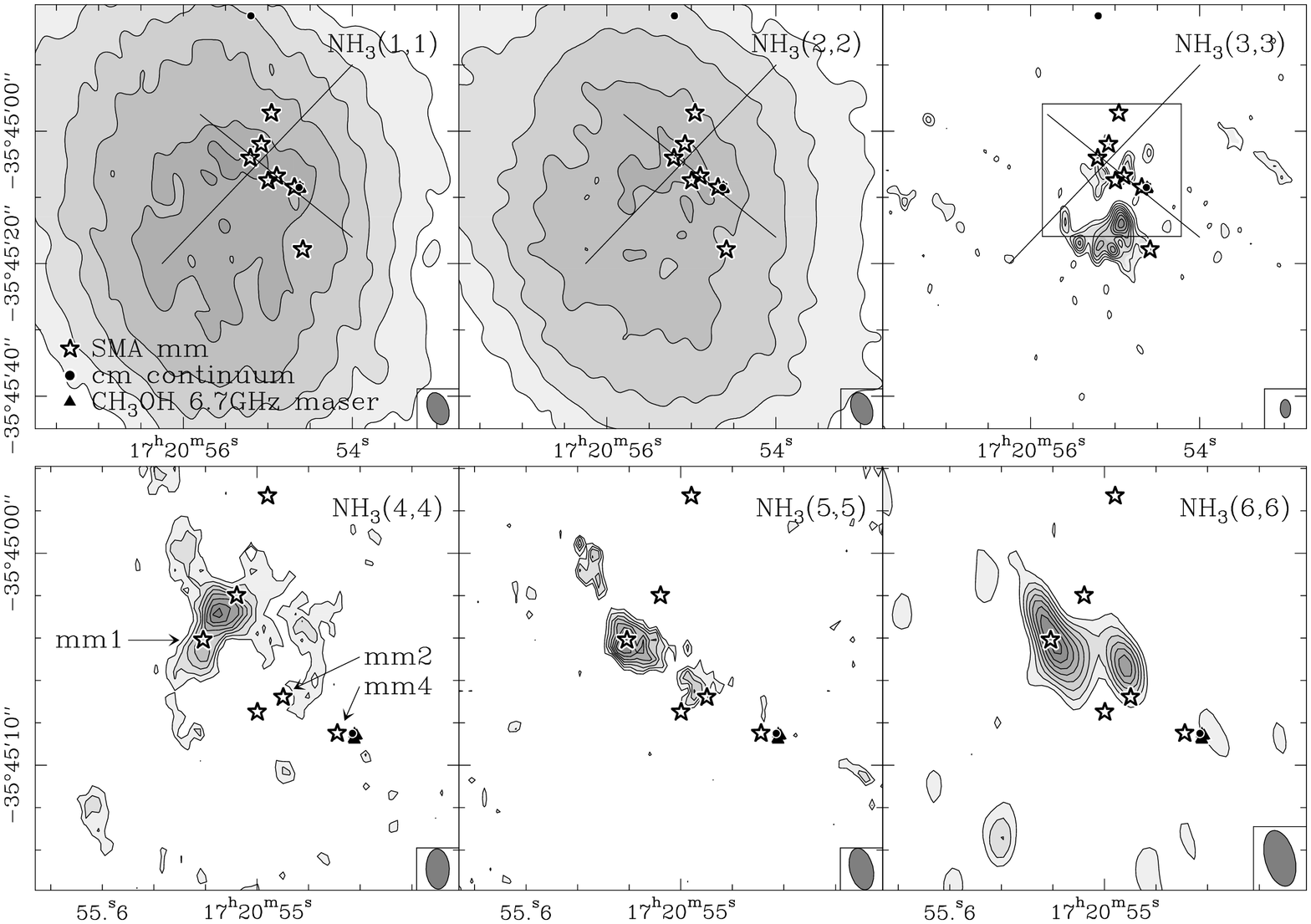}
\caption{Integrated main hyperfine component NH$_3$ inversion line
  images of the thermal emission toward NGC6334I(N) are shown in
  grey-scale with solid contours.  The (1,1) and (2,2) images are
  adapted from \citet{beuther2005e}. The top-row images are shown on
  larger spatial scales than the bottom-row images, the inlay size for
  the bottom-row images is shown in the (3,3) panel.  The contour
  levels are 58(58)348\,mJy/beam and 33(33)198\,mJy/beam for the
  NH$_3$(1,1) and (2,2) lines, respectively. The (3,3) to (6,6)
  integrated images are contoured from 25 to 95\% (step 10\%) of the
  peak emission with peak values of 12.9, 11.6, 10.0 and
  10.0\,mJy\,beam$^{-1}$, respectively. The markers are labeled in the
  top row, SMA mm continuum emission is from Hunter et al.~(subm.),
  CH$_3$OH class {\sc ii} maser emission from \citet{walsh1998} and
  the cm continuum emission from \citet{carral2002}. The mm continuum
  sources mm1, mm2 and mm4 from \citet{hunter2006} are labeled in the
  bottom-left panel. The large cross indicates the two main axis of
  the two identified molecular outflows. The synthesized beams are
  shown at the bottom-right of each panel. {\it The data to this
    Figure are also available in electronic form at the CDS via
    anonymous ftp to cdsarc.u-strasbg.fr (130.79.128.5) or via
    http://cdsweb.u-strasbg.fr/cgi-bin/qcat?J/A+A/.}}
\label{ngc6334in_images}
\end{figure*}

One of the surprising results of our previous NH$_3$(1,1), (2,2) and
CH$_3$OH (around 25\,GHz) study was the absence of compact molecular
line emission toward the main mm continuum peak
\citep{beuther2005e}. The NH$_3$ emission of the low excitation lines
peaks approximately at the mm continuum position, but it is extremely
extended without a clearly peaked morphology indicative of compact gas
cores (Fig.~\ref{ngc6334in_images} reproduces the data from
\citealt{beuther2005e}). Therefore, one of the aims of this new study
was to search for compact warm gas components associated with the mm
continuum sources, the strongest of which is the mm1 source labeled in
Fig.~\ref{ngc6334in_images}. The NH$_3$(3,3) emission appears less
extended than the (1,1) and (2,2) maps but it is still relatively
extended and mainly south of the mm continuum peaks. It is spatially
associated with neither of the two outflows but rather located between
their two southern outflow lobes. Perhaps there is some kind of
interaction zone between the two molecular outflows which
preferentially excites the NH$_3$(3,3) line. However, this is far from
clear and we refrain from further interpretation of the NH$_3$(3,3)
morphology here. Going to the higher excitation lines, the NH$_3$(4,4)
line is the first transition that shows compact emission close to the
mm continuum peaks. However, the emission peak is offset by about two
arcseconds to the north-west from the main mm continuum peak, mm1. This
may partially be an opacity effect because imaging the satellite
hyperfine component at negative velocities -- although the peak is
weak on a $5\sigma$ level -- it peaks toward the main mm source mm1.
Only the two highest excited lines, NH$_3$(5,5) and (6,6) clearly peak
toward the strongest mm peak mm1. Both maps show a secondary peak a
few arcseconds south-east associated with mm2.  We do not detect any
compact NH$_3$ emission toward mm4 which is associated with a cm
continuum source and a Class {\sc ii} CH$_3$OH maser
\citep{walsh1998,carral2002,hunter2006}.

\begin{figure}[htb]
\includegraphics[angle=-90,width=8.8cm]{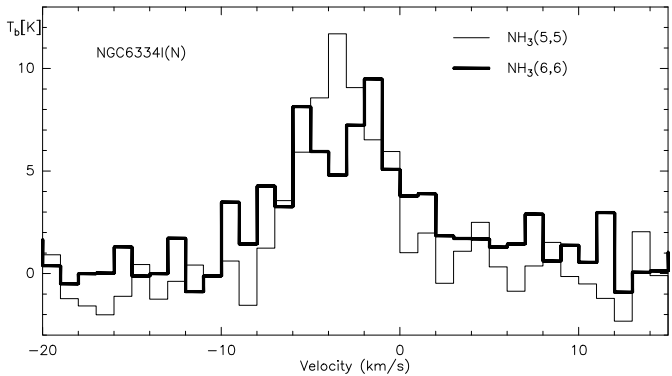}
\caption{NH$_3$(5,5) and (6,6) spectra in NGC6334I(N) extracted after
  imaging toward the their peak positions associated with the main mm
  continuum peak. The spectral resolution is 1\,km\,s$^{-1}$ and the spatial
  resolution $2.8\times2.4''$.}
\label{ngc6334in_disk}
\end{figure}

Only the NH$_3$(5,5) and (6,6) lines appear reasonable for a
temperature estimate toward the main mm continuum and high-excitation
NH$_3$ peak, because the NH$_3$(1,1) and (2,2) inversion lines trace
only extended gas, the (3,3) line shows the strange emission south of
the mm sources, and the (4,4) line peak is also offset from mm1.
Figure \ref{ngc6334in_disk} presents the NH$_3$(5,5) and (6,6) spectra
measured toward their main emission peaks toward mm1. While the
NH$_3$(5,5) line shows a relatively simple single-peaked profile, the
NH$_3$(6,6) line shows a double-peaked profile with the two emission
peaks separated by approximately $\pm 2$\,km\,s$^{-1}$ from the
peak-velocity derived from the NH$_3$(5,5) profile. With a $1\sigma$
rms of the NH$_3$(6,6) spectrum of 1.36\,K, the difference between the
emission peak at $-1.5$\,km\,s$^{-1}$ and the dip at
$-3.5$\,km\,s$^{-1}$ is $\sim 3.4\sigma$. It is not possible to derive
a 1 component fit for this (6,6) profile which could then be used with
the (5,5) spectrum to derive a temperature estimate similar to the
case in NGC6334I.  Therefore, we refrain from a rotational temperature
estimate here, however, kinetic temperatures in excess of 100\,K are
needed to explain the detection of the highly excited NH$_3$(5,5) and
(6,6) lines.

\section{Discussion}

\subsection{The outflow driving source in NGC6334I}
\label{outflowdriving}

What is the driving source of the large-scale north-east south-west
outflow in NGC6334I first reported by \citet{bachiller1990} and later
confirmed by \citet{kraemer1995} and \citet{hunter2006}? 

The larger line-widths observed toward mm1 compared to mm2 indicate
more internal motion toward mm1 -- either turbulent motion or
organized motion due to infall, outflow or rotation. Furthermore, the
fact that the highest excitation lines preferentially peak on the
strongest mm continuum source mm1 indicates that this is the warmest
source and potentially the strongest power house of the region, which
would favor that source as the driving source of the outflow. However,
mm line and continuum observations with the Submillimeter Array
(Hunter et al.~in prep.) show that mm2 has a higher line to continuum
ratio and that mm1 and mm2 are equally bright in several CH$_3$OH
lines with energy levels above ground of a few 100\,K. Therefore, we
cannot unambiguously conclude whether mm1 is the warmer source or not.
Furthermore, the NH$_3$(6,6) maser peak is clearly associated with the
mm continuum source mm2, and the position angle (PA) of the elongation
of the maser (6,6) feature is $\sim$48$^{\circ}$ from north,
comparable to the PA of the CO outflow of $\sim$46$^{\circ}$
\citep{bachiller1990}. This same orientation of the (6,6) maser
elongation with the large-scale molecular outflow indicates that the
maser emission may be produced by shock interaction of the outflow
with the ambient core gas close to the driving sources (in contrast to
the NH$_3$(3,3) maser features further outside). Therefore, we suggest
that this secondary mm source, mm2, may be the driving source of the
strong high-velocity molecular outflow.

\subsection{Double-peaked NH$_3$(6,6) emission in NGC6334I(N)}

The double-peaked NH$_3$(6,6) profile toward the main mm continuum
peak in NGC6334I(N) is peculiar and requires further consideration. In
typical infall studies, usually an optically thin and an optically
thick line are observed, and if the optically thick line shows an
absorption dip with a blue-red asymmetry at the velocity of the peak
emission of the optically thin line, this is a first indicator of
potential infall motion (e.g., \citealt{myers1996}). Fitting the
NH$_3$(5,5) spectrum shown in Figure \ref{ngc6334in_disk}, the optical
depth of the main hyperfine component is of order unity.  Although
NH$_3$(6,6) is part of the statistically favored ortho-NH$_3$ (its
statistical weight is double that of para-NH$_3$), and we cannot
precisely determine the optical depth of this higher excitation (6,6)
line from its double-peaked profile, it is unlikely to be very much
larger than that of the (5,5) line.  Therefore, a typical infall
interpretation appears unlikely.

Another possibility to produce such a spectrum is the presence of
multiple gas components at different velocities, perhaps from an
embedded binary system. In this case, the NH$_3$(6,6) line has to have
exactly the right optical depth to sample specifically these two gas
parcels.

Another way to get such a double-peaked profile is rotation of a
central circum-protostellar disk.  Such a double horn spectral profile
is expected if one observed such a disk in a near-edge-on
configuration where the gain path of the red-receding and
blue-approaching disk components are long enough to produce enough
emission at the corresponding velocities \citep{beckwith1993}. The
fact that we do not see such a double-peaked profile in the
NH$_3$(5,5) line may be explained by a larger optical depth of this
line which hence may not trace the inner disk structure at all.
Double-peaked profiles of some maser emission features were
interpreted in the past sometimes in the framework of rotation in a
Keplerian disk (e.g., \citealt{cesaroni1990,ponomarev1994}). However,
the emission observed in the NH$_3$(6,6) line toward NGC6334I(N) does
not appear maser-like and is likely of thermal nature.

With the current data, we cannot distinguish whether the double-peaked
spectrum is produced by multiple, isolated gas parcels or by a
contiguous structure like a rotating disk. Nevertheless, let us follow
the disk proposal for a moment: assuming equilibrium between the
centrifugal and gravitational force in an edge-on disk (inclination
angle $i=90^{\circ}$), the radius of such a disk would be:

$$ r {\rm{[AU]}} = \frac{M {\rm{[M_{\odot}]}}}{1.13\times 10^{-3}\,\Delta v^2 {\rm{[km\,s^{-1}]}}}\,sin^2i $$

With a separation $\Delta v$ between the two NH$_3$(6,6) peaks of
4\,km\,s$^{-1}$ and an approximate enclosed mass of 20\,M$_{\odot}$ estimated
from the mm continuum data \citep{hunter2006}, the approximate
diameter of this tentative disk-like structure is 2200\,AU,
corresponding to an angular size of $1.3''$. 

To discriminate between the different possibilities, higher spatial
resolution observations of the NH$_3$(6,6) line will be required to
resolve and image this potential massive accretion disk in more
detail. Furthermore, one can look for even higher excitation NH$_3$
lines: In the disk scenario, one would expect line profiles similar to
that observed in the (6,6) line, but with a larger velocity separation
since the warmer gas would be found at smaller disk radii in a
centrally heated Keplerian disk. In a multi-component scenario,
different line opacities could result in the higher energy transitions
sampling different gas components, which would produce different line
profiles for the higher transitions.

\section{Conclusions and Summary}

High spatial resolution imaging of the highly excited NH$_3$ inversion
lines (3,3) to (6,6) revealed warm gas components toward both of the
twin cores NGC6334I and I(N). This is particularly surprising for
NGC6334I(N) because this region was always considered to be the
prototypical massive cold core. While NGC6334I(N) is still in a very
young evolutionary stage it has already formed at least one central
massive protostar that has begun to heat up its surrounding
environment.  For the well-known hot core NGC6334I, the detection of
the higher excitation lines is less surprising. However, the optical
depths of the NH$_3$(1,1) to (4,4) are so high in both regions that
the hyperfine spectra are difficult to fit.  In NGC6334I, one can use
the (5,5) and (6,6) lines to estimate rotational temperatures toward
the mm continuum sources mm1 and mm2 of 86$\pm 20$ and 67$\pm
20$\,K, respectively.  However, since these two inversion lines are
from different NH$_3$ species (para- and ortho-NH$_3$), such
rotational temperature estimates have to been taken cautiously. We
also tried to estimate the opacities, and from that the rotational
temperatures and column densities, of the NH$_3$(3,3) and (4,4) lines
using the ratios between the main and satellite components of their
spectra. While these estimates qualitatively confirm the previous
estimates they also show the limitations and associated errors of the
NH$_3$ analysis in regions of that high column densities and
temperatures. Toward NGC6334I(N), a rotational temperature estimate is
not possible because the NH$_3$(6,6) spectrum is double-peaked and
hence does not allow reasonable line fits.  Therefore, we only
conclude that both regions harbor dense cores with kinetic
temperatures $>100$\,K. For better temperature estimates, one has
either to observe optically thin isotopologues like $^{15}$NH$_3$, or
go to even higher excitation inversion transitions like the
NH$_3$(7,7) and NH$_3$(8,8) lines, or try to model the spectra with
more sophisticated radiative transfer methods. A different approach is
to study other molecules with different abundances and energy levels
above ground. Many such molecules have a rich line spectrum in the
(sub)mm wavelength windows, and a complementary study with the
Submillimeter Array is currently in progress (Hunter et al.~in prep.).

We confirm the previous detection of the NH$_3$(3,3) maser toward the
two outflow lobes in NGC6334I. The north-eastern NH$_3$(3,3) maser
morphology even resembles the shape of typical outflow bow-shocks.
Furthermore, we report the first detection of NH$_3$(6,6) maser
emission toward the central cores in NGC6334I. The main NH$_3$(6,6)
maser peak is associated with the mm peak 2 and elongated along the
outflow axis. Therefore, we suggest that the strongest line and
continuum source in this region, mm1, is not the driving source of the
prominent molecular outflow, but rather the weaker sub-source, mm2.

While past observations revealed strong line emission on large-spatial
scales toward NGC6334I(N), no previous observations detected compact
gas cores at the center of the region. Observations of the highly
excited NH$_3$(4,4) to (6,6) lines for the first time detect compact
warm gas emission from the central main mm peak. The observed line
with the highest energy above ground, NH$_3$(6,6), is not
single-peaked toward that main mm continuum core, but it shows a
double-horn profile. With the current data we cannot differentiate
whether this double-horn line profile is caused by multiple gas
components along the line of sight (maybe even a binary system) or
whether it may be the signature of a potentially underlying massive
accretion disk. Higher spatial resolution observations of the
NH$_3$(6,6) or even higher excited lines that can resolve the
sub-structure of this core are required to solve this question.

\begin{acknowledgements} 
  We like to thank an anonymous referee as well as the editor Malcolm
  Walmsley for insightful comments clarifying this paper.
  H.B.~acknowledges financial support by the Emmy-Noether-Program of
  the Deutsche Forschungsgemeinschaft (DFG, grant BE2578).
\end{acknowledgements}

%\bibliography{/home/beuther/tex/bibliography}   
%\bibliography{/Users/henrikbeuther/paper/bibliography}
%\bibliographystyle{aa}    % this does the style, aa.bst necessary

\end{document}